\documentclass[reprint,prb]{revtex4-2}
\usepackage{graphicx,amssymb,amsfonts,amsmath}
\usepackage{color}
\usepackage{mathtools}
\usepackage{latexsym}
\usepackage{graphicx}
\usepackage{float}
\usepackage{dcolumn}
\usepackage{bm}
\usepackage{amssymb}
\usepackage{amsmath}
\usepackage{mathtools}
\usepackage[space]{grffile}
\usepackage{xcolor}
\usepackage{comment}
\usepackage{hyperref}
\usepackage{physics}
\usepackage{ulem}

\hypersetup{
    colorlinks=true,
    linkcolor=blue,
    filecolor=blue,      
    urlcolor=blue,
    citecolor=blue
    }

\usepackage{tikz, pgfplots,tikz-cd}
\usetikzlibrary{calc, decorations.markings, decorations.pathmorphing,shapes}
\tikzstyle string=[postaction={decorate},decoration={markings,
    mark=at position 0.5*\pgfdecoratedpathlength+.5*4pt with {\arrow[scale=1]{stealth[reversed]}}}]

\newcommand{\beq}{\begin{equation}\begin{aligned}}
\newcommand{\eeq}{\end{aligned}\end{equation}}

\begin{document}

\title{Non-invertible symmetries out of equilibrium: Eigenstate order and Floquet physics}
\author{Yabo Li}
\author{Aditi Mitra}
\affiliation{
Center for Quantum Phenomena, Department of Physics,
New York University, 726 Broadway, New York, New York, 10003, USA
}

\begin{abstract}
Through the study of the Rep($D_8$) non-invertible symmetry, we show how non-invertible symmetries manifest in dynamics. Results are presented for dynamics generated by Hamiltonians as well as Floquet unitaries. For both examples, the role of the non-invertible symmetry is studied through the appearance of non-invertible symmetry protected edge modes. In addition, the role of the non-invertible symmetry for the Hamiltonian is studied through eigenstate order.
In particular, by considering the effect of symmetry preserving disorder, the non-invertible symmetry is shown to give rise to degeneracies in the spectra of the Hamiltonian that  can only be completely lifted at orders of perturbation that scale with system size. The eigenstates of disordered Hamiltonians, whose ground states correspond to non-trivial symmetry protected topological (SPT) states, are shown to have either trivial or non-trivial SPT order that are detected as non-zero expectation value of string order-parameters. In contrast, non-trivial SPT order is absent in the eigenstates of trivial SPT Hamiltonians with disorder. The interface between two different SPT phases host edge modes whose dynamics is studied numerically and analytically. The edge mode is shown to oscillate at frequencies related to different effective chain lengths that are weighted by the temperature, becoming an exact zero mode in the limit of zero temperature.  A Floquet model with the non-invertible symmetry is constructed whose edge mode is shown to exhibit period-doubled dynamics at low effective-temperatures. The zero and period-doubled edge modes differ from those in conventional SPTs by being  symmetric under the invertible symmetry, while being charged under the non-invertible symmetry.
\end{abstract}
\maketitle

\section{ Introduction} 
With the interest in generalized symmetries \cite{mcgreevy2023generalized}, one now recognizes that the Kramers-Wannier duality encountered in Ising models \cite{Kramers41}, is just one in a large family of generalized symmetries, known as non-invertible symmetries \cite{Fendley20,Fendley24,schafer2024ictp,shao2023s}. The term non-invertible captures the fact that the action of the symmetry and its conjugate, does not give unity. Most studies have explored the effect of non-invertible symmetries in the ground state of Hamiltonians. Here we explore whether these symmetries show any signatures far out of equilibrium, when the entire spectrum 
%of the Hamiltonian or Floquet unitary 
is important for the dynamics. We study both energy conserving non-equilibrium dynamics, i.e, dynamics generated by a Hamiltonian, as well as discrete time stroboscopic dynamics i.e,  dynamics generated by a Floquet unitary. For the latter, while energy is not conserved, a quasi-energy can still be defined \cite{Sambe73}.

Formally, non-invertible symmetries are described by fusion categories \cite{etingof2005fusion,thorngren2024fusion,inamura2021topological}.
In short, a fusion category $\mathcal{D}$ is specified by a set of simple objects $\{a\}$ with fusion rules
\beq
    a\cdot b = \sum_{c}N^{ab}_c c,
\eeq
where $N^{ab}_c$ are non-negative integers. The fusion of simple objects satisfies associativity up to an $F$-matrix, and these obey the pentagon identities for consistency.
For instance, the symmetry of the two dimensional (2D) Ising conformal field theory is given by the Ising category, which includes an ordinary $\mathbb{Z}_2$ symmetry $\eta$ and the non-invertible Kramers-Wannier symmetry $KW$, satisfying the fusion rule:
\beq KW^2 = 1+\eta. \eeq
Here, the symmetry is non-invertible because the action of Kramers-Wannier and its conjugate gives a projector to $\mathbb{Z}_2$ symmetry sectors. The non-invertibility arising due to the projection to symmetry sectors is a general property of the Tambara-Yamagami category \cite{tambara2000representations,bhardwaj2018finite,thorngren2024fusion}, of which the Ising category is an example.

Despite the Kramers-Wannier duality being the most well known example of a non-invertible symmetry, it is in fact not the simplest example because it is not anomaly-free \cite{chang2019topological,seiberg2024non}, implying that the Hamiltonian does not have a gapped symmetry preserving ground state, but rather is either gapless or the ground state breaks the symmetry. Formally, when a fusion category admits a fiber functor~\cite{tambara2000representations,thorngren2024fusion}, there is a corresponding non-invertible symmetry protected topological (SPT) phase, i.e., a non-degenerate short ranged entangled (SRE) state symmetric under the non-invertible symmetry, generalizing the notion of SPT phases with ordinary symmetry \cite{chen2011classification,turner2011topological,schuch2011classifying,lu2012theory,chen2013symmetry,chen2014symmetry,senthil2015symmetry}. One of the simplest such anomaly-free non-invertible symmetries is Rep$(D_8)$. 

In our work, we will discuss the charge under the action of non-invertible symmetries. For invertible symmetries, the charges are one-dimensional representations of the symmetries. Moreover, for a large class of invertible SPT phases, the charges decorate the domain walls~\cite{chen2013symmetry}, leading to the definition of string order parameters~\cite{den1989preroughening,kennedy1992hidden}. In this work, we will present a domain wall charge decoration picture of Rep$(D_8)$ non-invertible SPT phases, and study the corresponding string order parameters~\cite{lu2025strange}. These charges turn out to be one-dimensional representations of the non-invertible Rep$(D_8)$ symmetry. However, there is a different notion of charges corresponding to Thouless pumps. For invertible symmetries, the charges, i.e., the one-dimensional representations, are pumped through adiabatic cycles of one-dimensional SPT states~\cite{kitaev2011SRE,kitaev2013SRE,bachmann2024classification}. The classification of Thouless pumps for non-invertible symmetries is studied in Ref.~\cite{inamura20241+,li2025classification}, where it is shown that the pumped charges differ in different SPT phases of the same symmetry, and that they are not necessarily one-dimensional representations of the symmetry. The precise connection between these two interpretations of charges of non-invertible symmetries, one in the connection of string-order parameters, and the other in connection to Thouless pumps, is not understood yet, and will not be addressed in this work.

In what follows, we study the effect of Rep$(D_8)$ symmetry beyond the ground state sector. The time evolution of any observable now involves averaging over the entire spectrum. Thus,  out of equilibrium phenomena is also probed by studying properties of the spectra such as robustness of degeneracies and eigenstate order. These quantities signal the violation of the eigenstate thermalization hypothesis, and can lead to non-ergodic behavior in the autocorrelation functions, and the emergence of localized edge modes.

The paper is organized as follows. In Section \ref{model} we present the Rep$(D_8)$ category on a lattice, presenting examples of Hamiltonians whose ground states are trivial and non-trival SPTs of the Rep$(D_8)$ symmetry. In Section \ref{eg-order} we move away from  the ground state sector, presenting results for robust degeneracies and eigenstate order in Hamiltonians with quenched disorder. In Section \ref{edge} we study disorder free Hamiltonians and Floquet models with spatial boundaries, deriving results for zero and period doubled edge-modes, and discussing their stability away from the low (quasi)-energy sector. In Section \ref{concl} we present our conclusions. Details of derivations are relegated to four appendices. 

\section{Model} \label{model}
There are three distinct SPT phases corresponding to the three fiber functors of the Rep$(D_8)$ category \cite{thorngren2024fusion,seifnashri2024cluster}. The fixed-point Hamiltonians and ground state phases with this symmetry, along with other non-invertible symmetries, and for different lattice realizations were recently studied \cite{inamura2022lattice,seifnashri2024cluster,meng2024non,warman2024categorical,bhardwaj2024lattice,jia2024generalized,cao2024generating,li2024non,aksoy2025phases}. In this paper we investigate the dynamics generated by both Hamiltonians as well as  Floquet unitaries which have the Rep($D_8$) non-invertible symmetry. 

We study a lattice  realization of the Rep($D_8$) symmetry corresponding to a one-dimensional lattice of $L=4N$, $N\in$ {\rm int} sites, with one qubit on each site. The Rep($D_8$) symmetry is given by
\beq
    \eta_e=\prod_i X_{2i},\quad  \eta_o=\prod_i X_{2i+1}, \quad KT,
    \label{eq:symmetry operators}
\eeq
with the fusion relations 
\begin{subequations}
\begin{align}
    \eta_e^2=\eta_o^2=1,\,\, \eta_{e/o} KT= KT\eta_{e/o}=KT,\\
    KT^2=1+\eta_e+\eta_o+\eta_e\eta_o.
\end{align}
\end{subequations}
Above $KT$ is the non-invertible symmetry known as the Kennedy-Tasaki duality~\cite{kennedy1992hidden,li2023non}. It satisfies
\begin{subequations}
\begin{align}
    KT X_i = X_i KT,\\ 
    KT Z_{i-1}Z_{i+1} = Z_{i-1}X_{i}Z_{i+1} KT.
\end{align}
\end{subequations}

Two equivalent ways to implement this duality are through a sequential circuit construction~\cite{seifnashri2024cluster} and a matrix product operator (MPO) construction~\cite{meng2024non} (see Appendix \ref{appA}). For the former,the $KT$ duality operator can be expressed as
    $KT = T D_e D_o$, where $T$ performs one-site lattice translation, while $D_{e,o}$
are sequential two and three qubit gates, see Appendix \ref{appA}. We note that even though a lattice translation appears explicitly in the above definition of the $KT$ operator, the local operator mapping given by $KT$ does not mix with translation. This is in contrast to the Kramers-Wannier duality in the transverse field Ising model, 
\beq
    KW X_i = Z_i Z_{i+1} KW,\\ 
    KW Z_{i}Z_{i+1} = X_{i+1} KW.
\eeq
The above equations shows that $KW^2$ maps $X_i$ to $X_{i+1}$, implying that $KW$ mixes with lattice translation.
The MPO construction of $KT$ is (see Appendix \ref{appC})
\begin{align}
 KT&= \sum_{i_1,j_1,\cdots,i_L,j_L}\Tr(A^{i_i}_{j_1} A^{i_2}_{j_2} \cdots A^{i_L}_{j_L})\nonumber\\
 &\times \ket{i_1,i_2,\cdots, i_L}\bra{j_1,j_2,\cdots,j_L},
    %KT= \Tr(A^1 A^2 \cdots A^L),
\end{align}
where $A$ is the following bond dimension-two matrix
\beq
    A^0_0=A^1_1=\ket{+}\bra{0},\quad A^0_1=A^1_0=\ket{-}\bra{1}.
    \label{eq:KT-MPO}
\eeq
We note that an alternate realization of the Rep($D_8$) symmetry on the lattice is \cite{seifnashri2024cluster}  $\eta_e,\eta_o,D$, with $D$ and $KT$ being related by a  conjugation by control-Z gates: $\prod_i CZ_{i,i+1}$.
In Ref.~\cite{meng2024non,warman2024categorical}, the Rep($D_8$) symmetry was realized on a $D_8$ qudit chain and a three-qubit spin chain using MPOs,  and the MPO construction in \eqref{eq:KT-MPO} can be reduced from it.

Some local operators that are symmetric under Rep$(D_8)$ symmetry are 
\begin{align}
  &  X_i,\ Z_{i-1}Z_{i+1}(1+X_{i}),\ Z_{i-1}Z_{i+3}(1+X_{i}X_{i+2}),
  \nonumber\\
  &Z_{i-1}Z_{i+5}(1+X_{i}X_{i+2}X_{i+4}),\ \cdots\nonumber.
\end{align}
These will be employed to construct Hamiltonians and Floquet unitaries.

{\bf Rep($D_8$) SPT phases:} SPT phases are the unique, gapped ground states of Hamiltonians formed by local symmetric operators. Furthermore, the fixed-point Hamiltonians of SPT phases are usually formed by commuting projectors since the ground state has vanishing correlation length. 
There are only three distinct SPT phases with Rep($D_8$) symmetry \cite{thorngren2024fusion}, with lattice models for these three SPT phases given in Ref.~\cite{seifnashri2024cluster}, and dubbed the trivial, and the non-trivial odd, even SPTs. Indeed, the three Rep$(D_8)$ SPT phases are the ground states of \footnote{The three Rep($D_8$) SPT models we study are related to the SPT models in Ref.~\cite{seifnashri2024cluster} by a conjugation with $\prod_i CZ_{i,i+1}$},
\begin{subequations}
\begin{align}
    &H_{\rm{trivial}}
    =-\sum_i X_i, \label{eq:trivial}\\
    &H_{\rm{odd}}=\sum_{i}X_{2i}
  + \sum_{i}Z_{2i-1}X_{2i+1}Z_{2i+3}\biggl(\frac{1+X_{2i}X_{2i+2}}{2}\biggr), \label{eq:odd}\\
&H_{\rm even}=\!\!\sum_{i}Z_{2i-2}X_{2i}Z_{2i+2}\biggl(\frac{1+X_{2i-1}X_{2i+1}}{2}\biggr)
\!\!\!+\!\!\sum_{i}X_{2i+1}.
    \label{eq:even}
    \end{align}
\end{subequations}
We use periodic boundary conditions for models throughout this work, unless specified otherwise. The ground states of the above three models cannot be related by a finite-depth circuit of symmetric gates. Furthermore, at the interface of two different SPT ground-states, there are edge modes protected by the Rep$(D_8)$ symmetry. In Appendix \ref{appB}, by constructing finite depth circuits of symmetric gates, we establish equivalences between the ground states of Hamiltonians comprising of different combination of local operators. 

\section{ Non-invertible SPT eigenstate order} \label{eg-order}
Focusing on the non-invertible Rep($D_8$) SPT phases with quenched disorder, we now present evidence for non-invertible eigenstate order. We employ the same metrics employed in studies of localization-protected eigenstate order for ordinary  symmetries \cite{Bauer2013area,huse2013localization,Chandran2014many,bahri2015localization,Parameswaran2018many}, such as non-decaying expectation value of string order-parameters for typical eigenstates \cite{Chandran2014many}, and symmetry-protected boundary modes for eigenstates~\cite{bahri2015localization}. We uncover important differences between ordinary and non-invertible symmetries. 

To begin with, we notice that the even and odd SPT Hamiltonians in \eqref{eq:odd},\eqref{eq:even} have different spectra statistics from the trivial SPT Hamiltonian in \eqref{eq:trivial}. For the trivial Hamiltonian, there are some accidental degeneracies in the spectrum. For example, denoting $X_i|+\rangle = |+\rangle$, the excited states $Z_{i}\ket{+}^{\otimes L}$ for all sites $i$ have the same energy. A quenched disorder could break all these degeneracies and lead to the localization of eigenstates. However, for the even and odd  SPT Hamiltonians, some degeneracies are protected by the Rep$(D_8)$ symmetry. Let us take $H_{\rm odd}$ as an example. Whenever $X_{2i}X_{2i+2}=-1$, there is a degeneracy in the spectrum, due to the vanishing of the corresponding term (second term in \eqref{eq:odd}) in the Hamiltonian, and quenched disorder in the Hamiltonian will not break these degeneracies at low orders in perturbation theory. 

To be more explicit, we consider the following Hamiltonian composed of terms of the trivial and the odd SPT Hamiltonians,
\begin{align}
&H_d= H_{\rm odd} + H_{\rm trivial}\nonumber\\
& = 
    \sum_{i}J_{2i+1}Z_{2i-1}X_{2i+1}Z_{2i+3}\biggl(\frac{1+X_{2i}X_{2i+2}}{2}\biggr)\nonumber\\
    &+\sum_{i}J_{2i}X_{2i} - \sum_i B_i X_i.
    \label{eq:disorder_oddtrivial}
\end{align}
In the clean case where the coupling strengths $\{J_{i}\}$ and $\{B_i\}$ are uniform, the above Hamiltonian is in a gapped SPT phase when $|J| < |B|$, and in another gapped SPT phase when $|J| > |B|$. At $|J| = |B|$ a phase transition occurs. When $J= -B$, the effective Hamiltonian for the low energy sector is $H=-B\sum_i Z_{2i-1}X_{2i+1}Z_{2i+3} -B\sum_i X_{2i-1} -2B\sum_i X_{2i}$ where $X_{2i}$ are fixed to $+1$ by the last term. Hence the critical point describes the transition between the cluster phase and the paramagnetic phase, which is a $c=1$ conformal field theory~\cite{wolf2006quantum}. When $J= B$, the terms $J X_{2i}$ and $-BX_{2i}$ cancel and the $X_{2i}$'s are not fixed in the low energy sector. Depending on the configuration $\{X_{2i}\}$, the Hamiltonian describes a superposition of critical cluster-paramagnetic models on open chains with various length. The lowest energy sector is given by the critical chain of the longest length, hence it is still described by a $c=1$ conformal field theory. In order to extract the eigenstate order, one needs to lift the accidental spectral degeneracies, hence we consider a quenched disorder on all coupling strengths, with $J_i\in [-W_J, W_J]$ and $B_i\in [-W_B, W_B]$.

\begin{figure*}[t!]
        % \centering
        \includegraphics[width=0.4\textwidth]{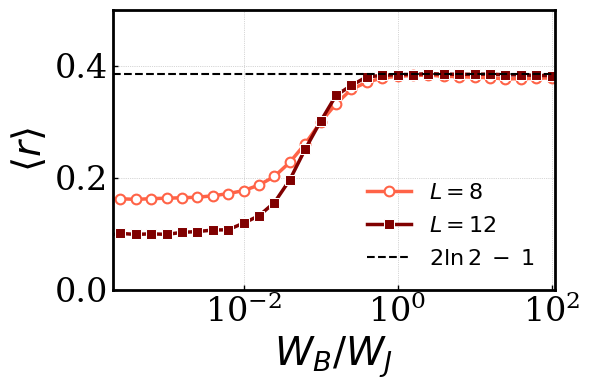}
         \includegraphics[width=0.4\textwidth]{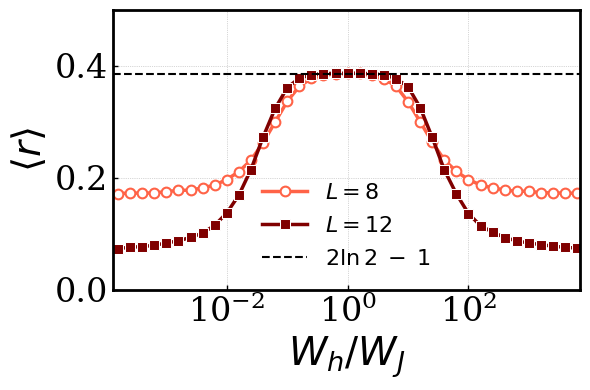}
         \caption{The $\langle r\rangle$-ratio as a function of $W_B/W_J$ for 
         $H_{\rm odd}(\{J_i\}) + H_{\rm trivial}(\{B_i\})$ (left panel), $H_{
        \rm odd}(\{J_i\}) + H_{\rm even}(\{h_i\})$ (right panel). The dashed line corresponds to the 
        $\langle r\rangle$-ratio for Poisson distributed levels. 
        Each data point corresponds to averaging over at least $1000$ disorder realizations. The $\langle r\rangle$-ratio in two graphs when $W_{B(h)}/W_J\approx 10^{-4}$ are noticeably different. This is because of the distinct nature of the perturbations that lift the degeneracies.}
        \label{fig:rratio}
\end{figure*}

In Fig.~\ref{fig:rratio}, we show the $\langle r\rangle$-ratio defined as $\langle r\rangle = \langle {\rm min}\left[\Delta_{i},\Delta_{i+1}\right]/{\rm max}\left[\Delta_{i},\Delta_{i+1}\right]\rangle$, $\Delta_i = |E_{i+1}-E_i|$, 
where $\{E_i\}$ are sequentially ordered energy levels, and the bracket denotes an average over the entire spectrum and over disorder realizations \cite{Oganesyan07localization,Atas2013distribution}. A small value for the $\langle r\rangle$-ratio implies degeneracies in the spectra. When $W_B \gg W_J$ (left panel), we are in the perturbative regime of the trivial (paramagnetic) Hamiltonian, which clearly does not have robust degeneracies. Due to the projectors, this model appears to be an interacting fermion problem after fermionization, and so it is a bit surprising that the $\langle r\rangle$-ratio agrees well with the Poisson expectation ($2\ln 2 -1$) \cite{Oganesyan07localization,Atas2013distribution}. The reason for this is that, the model is still exactly solvable because all $X_{2j}$ operators are good quantum numbers, and we can decompose the Hilbert space into different $\{s_{2j}\}$ sectors where $s_{2j}=\pm 1$ are the eigenvalues of $X_{2j}$. Within each sector, the Hamiltonian is composed of just $Z_{2i-1}X_{2i+1}Z_{2i+3}$ and $X_{2i+1}$, i.e, it is a cluster model on odd sites. This model can be mapped to fermion bilinears upon fermionization. Hence, with quenched disorder, each sector experiences Anderson localization. 

Although the model is exactly solvable in sectors, a qualitative difference in the spectrum is visible when $W_B/W_J \gg 1$ vs $W_B/W_J \ll 1$. The low plateau in the case of $W_B/W_J \ll 1$ indicates a spectral degeneracy robust against the perturbations corresponding to disordered $\{B_i\}$ couplings. This is attributed to the KT symmetry. We explain this below.

When $W_B/W_J \ll 1$, we are in the perturbative regime of the odd SPT Hamiltonian, and the $\langle r\rangle$-ratio saturates to a smaller value for fixed system size $L$. When $W_J$ is strong, we can see from perturbation theory that the degeneracies persist. We denote the eigenstates of $H_{\rm odd}$ as $\ket{\{s_i\}}$, where $s_i=\pm 1$ means that the state is stabilized by
\beq
    s_{2j}X_{2j}, \quad s_{2j+1}Z_{2j-1}X_{2j+1}Z_{2j+3}.
\eeq
The energy is given by
\beq
    E_{\{s_i\}} = \sum_{i} s_{2i} J_{2i} + s_{2i+1}J_{2i+1}\frac{1+s_{2i}s_{2i+2}}{2}.
\eeq
Under $KT$, the stabilizers are modified as follows $KT s_{2j} X_{2j}= s_{2j} X_{2j}KT $ and 
$KT s_{2j+1}Z_{2j-1}X_{2j+1}Z_{2j+3} = s_{2j+1}Z_{2j-1}X_{2j+1}Z_{2j+3} X_{2j} X_{2j+2}KT $. Thus,
under $KT$ the  state $\ket{\{s_i\}}$ is transformed to one that is stabilized by $s_{2j} X_{2j},s_{2j+1}Z_{2j-1}X_{2j+1}Z_{2j+3} s_{2j} s_{2j+2} $, i.e, $KT$ maps $\ket{\{s_i\}}$ to 
\beq
    \ket{\{s_i\}} \xrightarrow{KT} 
    \prod_j \biggl(Z_{2j-1}Z_{2j+1}\frac{1-X_{2j}}{2} + \frac{1+X_{2j}}{2}\biggr)\ket{\{s_i\}},
    \label{eq:KT-eigenstates}
\eeq
if it is in the $\mathbb{Z}_2^2$ symmetric sector, otherwise, the state is eliminated by $KT$. To verify, we note that the unitary $U = \prod_j (Z_{2j-1}Z_{2j+1}\frac{1-X_{2j}}{2} + \frac{1+X_{2j}}{2})$ in the above equation satisfies
\beq
    X_{2j} U &= U X_{2j}, \\
    Z_{2j-1}X_{2j+1}Z_{2j+3}  U &= U Z_{2j-1}X_{2j+1}Z_{2j+3} \cdot X_{2j} X_{2j+2}.
\eeq
The state $U \ket{\{s_i\}}$ is thus stabilized by $s_{2j} X_{2j}$ and $s_{2j}s_{2j+1} s_{2j+2}Z_{2j-1}X_{2j+1}Z_{2j+3}$ because
\beq
    &s_{2j} X_{2j} \Big(U \ket{\{s_i\}}\Big) = U \Big(s_{2j} X_{2j} \ket{\{s_i\}}\Big) = U \ket{\{s_i\}},
\eeq
and
\beq
    &s_{2j} s_{2j+1}s_{2j+2}Z_{2j-1}X_{2j+1}Z_{2j+3}   \Big(U \ket{\{s_i\}}\Big)\\
    & = U \Big(s_{2j+1}Z_{2j-1}X_{2j+1}Z_{2j+3}\Big) \\
    &\quad \quad \quad \cdot\Big(s_{2j}X_{2j}\Big)\Big(s_{2j+2}X_{2j+2}\Big)\ket{\{s_i\}}\\
    &= U \ket{\{s_i\}}.\\
\eeq

In perturbation theory, the energy correction at any order can always be written as a function of 
$    V_{\{s_i\},\{s'_i\}}\equiv \bra{\{s_i\}}V\ket{\{s'_i\}}$, $
    E_{\{s_i\},\{s'_i\}}\equiv E_{\{s_i\}}-E_{\{s'_i\}}$. 
For example, at  second order,
$    E_{\{s_i\}}^{(2)}=\sum_{\{s_i'\}\neq\{s_i\}}|V_{\{s_i\},\{s'_i\}}|^2/E_{\{s_i\},\{s'_i\}}$.
For a quenched disordered odd SPT Hamiltonian, when a state $\ket{\{s_i\}}$ is not invariant under $KT$, the state 
$    \ket{\{\tilde{s}_i\}}\propto KT \ket{\{s_i\}}$
has the same energy. Thus for any $V_{\{s_i\},\{s'_i\}}$ and $E_{\{s_i\},\{s'_i\}}$, there is always a companion after $KT$ (i.e., $V_{\{\tilde{s}_i\},\{\tilde{s}'_i\}}$ and $E_{\{\tilde{s}_i\},\{\tilde{s}'_i\}}$) of the same value. As a result, the energy correction for the two degenerate states $\ket{\{s_i\}}$ and $\ket{\{\tilde{s}_i\}}$ are always the same. Hence, the only way to lift the degeneracy between them is for the perturbation to form an off-diagonal contribution in the effective Hamiltonian in the degenerate subspace. Since the action of $KT$ might involve products of extensively many symmetric local operators as shown in \eqref{eq:KT-eigenstates}, the degeneracy will be completely lifted only to an extensively large power of the perturbation.

To further highlight the degeneracies due to the non-invertible symmetry, we now consider the following Hamiltonian composed of terms in both the odd and the even SPT Hamiltonians
\beq
    H_d' &= H_{\rm odd} + H_{\rm even} =\sum_{i}J_{2i}X_{2i}+\sum_i h_{2i+1} X_{2i+1}\nonumber\\
    &+\sum_{i}J_{2i+1}Z_{2i-1}X_{2i+1}Z_{2i+3}\biggl(\frac{1+X_{2i}X_{2i+2}}{2}\biggr)  \nonumber\\
    &+\sum_{i}h_{2i}Z_{2i-2}X_{2i}Z_{2i+2}\biggl(\frac{1+X_{2i-1}X_{2i+1}}{2}\biggr),
    \label{eq:disorder_oddeven}
\eeq
where all coupling strengths are disordered, with $J_i\in [-W_J, W_J]$ and $h_i\in [-W_h, W_h]$. This Hamiltonian non-trivially couples the even and odd numbered sites, and does not appear to be exactly-solvable. Nevertheless, when $W_J \gg W_h$ or $W_h \gg W_J$, from the perturbation argument given above, one expects the $\langle r\rangle$-ratio to acquire a small value that decays with system size. This is also consistent with the numerics, see right panel of Fig.~\ref{fig:rratio}. When $W_J \approx W_h$, the $\langle r\rangle$-ratio approaches the Poisson expectation value, suggesting localization.  
\begin{figure*}[t!]
     \centering
         \includegraphics[width=0.4\textwidth]{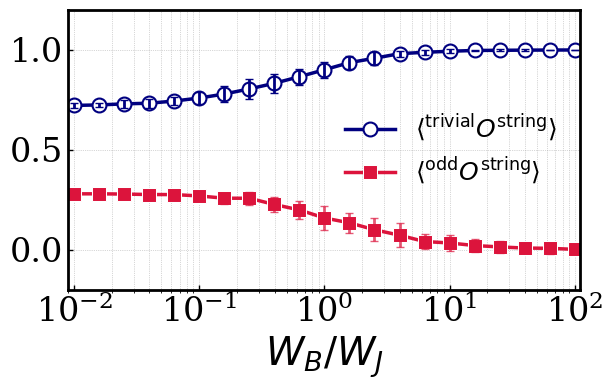}
         \includegraphics[width=0.4\textwidth]{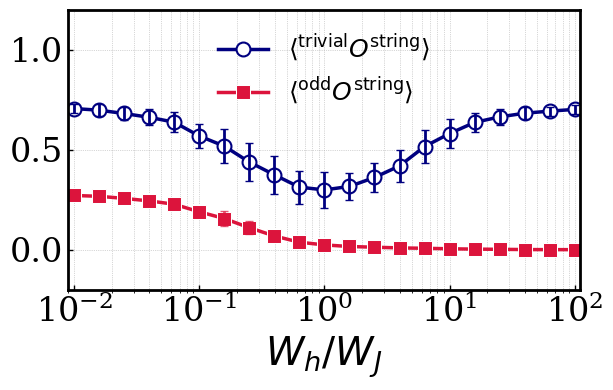}
        \caption{The average of odd and trivial string order parameter $(^{\rm odd/trivial}\mathcal{O}^{\rm string})_{i,i;I,J}$ with $I=1,J=2,$ and system size $L=12$ for $H_{\rm odd}(\{J_i\}) + H_{\rm trivial}(\{B_i\})$ (left panel), and  $H_{
        \rm odd}(\{J_i\}) + H_{\rm even}(\{h_i\})$ (right panel). Each data point corresponds to averaging over at least $100$ disorder realizations. Vertical lines denote the variance in the spectral average from different disorder realizations.
        On the right-panel (not shown), $(^{\rm even}\mathcal{O}^{\rm string})_{i,i;I,J}$ undergoes a transition from zero to non-zero as $W_h/W_J$ increases.}
        \label{fig:string order parameter}
\end{figure*}

Having established degeneracies in the spectra due to $KT$ symmetry, we now present the analog of  eigenstate order for the non-invertible symmetry. The string order parameters for Rep($D_8$) SPT states \cite{lu2025strange} can be expressed as MPOs that involve truncating the $KT$ operator \eqref{eq:KT-MPO} at sites $I,J$, and introducing boundary operators (see Appendix \ref{appC}). We define,
\begin{align}
    &(^{\text{trivial}}\mathcal{O}^{\text{string}}_{KT})_{i,j; I, J} = \bra{i}A_{2I}\cdots A_{2J+1}\ket{j},\\
 &(^{\text{odd}}\mathcal{O}^{\text{string}}_{KT})_{i,j; I, J} = Z_{2I-3}^{i}X_{2I-1}^{i}Z_{2I-1}^{1-i}\nonumber\\
 &\times \bra{i}A_{2I}\cdots A_{2J+1}\ket{j}
 Z_{2J+3}^{j}X_{2J+3}^{1-j}Z_{2J+5}^{1-j}.   
\end{align}
These string order parameters are $2\times 2$ matrix valued operators with indices $i,j$. For the ground states of the odd SPTs, the expectation value of the odd-SPT string order parameter is non-zero at long distance $|I-J|\gg 1$, with a spectral norm $||\langle (^{\text{odd}}\mathcal{O}^{\text{string}}_{KT})_{I, J} \rangle||_2=1$. The norm of $B$ is defined as 
$
||B||_2\equiv \sqrt{\lambda_{\text{max}}(B^* B)}$, $\lambda_{\rm max}(B^*B)$ being the largest eigenvalue of $B^*B$.

In contrast,
in the ground states of the trivial SPTs, the expectation value of this operator vanishes for $I\neq J$, while the trivial string order is non-zero and  long-ranged, i.e, $
    ||\langle (^{\text{trivial}}\mathcal{O}^{\text{string}}_{KT})_{I, J} \rangle||_2 \equiv 1$.
Therefore, the non-decaying behavior of these string order parameters are diagnostics for the non-invertible SPT states.

The eigenstate order of an ordinary SPT is such that the average of the corresponding string order parameter, over the spectrum is non-zero, while the average of the string order parameter for other phases  vanishes~\cite{Chandran2014many,long2023phenomenology}. Here we show that the non-invertible SPT eigenstate orders are different. In Fig.~\ref{fig:string order parameter}, we present the average of the expectation of the string order parameters for trivial and the non-trivial SPTs. We find  that when the system is slightly perturbed from $H_{\rm odd}$ either by $H_{\rm trivial}$ or $H_{\rm even}$, the expectations of both the trivial as well as the odd-SPT string order-parameters are non-zero. Hence, the spectrum always contains eigenstates with both odd SPT order and trivial SPT order. As we tune the disorder strength such that the Hamiltonian enters the perturbative regime of $H_{\rm trivial}$, only the trivial string order parameter is non-zero. Thus the entire spectrum contains only trivial SPT order, similar to the situation for SPT eigenstate order with ordinary symmetry. On the other hand, when considering $H_{\rm odd} + H_{\rm even}$, and when turning  on the disorder strength in $H_{\rm even}$, we have a transition from the eigenstate order of the odd SPT to the even SPT. In the last section we saw that during the transition, the $\langle r \rangle$-ratio increases and approaches the Poisson expectation. Here, we can also see that $\langle^{\rm odd}\mathcal{O}^{\rm string}_{KT}\rangle$ goes to zero during the transition. Since $H_{\rm odd}$ and $H_{\rm even}$ differ by a lattice translation, so do their string order parameters. Thus, $\langle^{\rm even}\mathcal{O}^{\rm string}_{KT}\rangle$ increases to non-zero values after the transition.

As we end this section on eigenstate order we comment that the study of non-invertible eigenstate order in Floquet systems with quenched disorder is left for future studies. Instead, in the remaining paper, we study dynamics of disorder-free systems with interfaces. 

\section{ Edge mode dynamics of Rep($D_8$) symmetric models} \label{edge}
We now consider dynamics generated both by a Hamiltonian as well as a Floquet unitary, and for a system with boundaries. For ordinary symmetries, eigenstate SPT order implies that symmetry-protected edge degeneracies persist not only in the ground state but throughout the entire spectrum. In Hamiltonian dynamics, this manifests as the existence of a localized zero mode. Moreover, such localized zero modes are not restricted to Hamiltonian dynamics with eigenstate order, but can also arise in more general time-periodic or Floquet unitary evolutions. The Floquet SPT phases for ordinary symmetries  can be represented by a zero-correlation length (fixed-point) Floquet unitaries that are closely related to the fixed-point SPT Hamiltonians with the same ordinary symmetry, and enriched by the discrete time translation symmetry \cite{potter2016classification,von2016phase}. The fixed-point unitaries are usually non-interacting, with single-particle quasi-spectra. Different classes of Floquet unitaries are protected by the symmetry in that going from one class to another would encounter a single-particle quasi-spectral gap closing. Some properties of the fixed-point unitaries persist under finite perturbation and can be detected via localized  edge modes that are stable for long times \cite{potirniche2017floquet,bomantara2018quantum,yates2019almost,Zhang2022Digital,samanta2025}. We plan to study the analogous physics, but with the non-invertible symmetry.

\begin{figure}[t!]
        % \centering
        \includegraphics[width=0.35\textwidth]{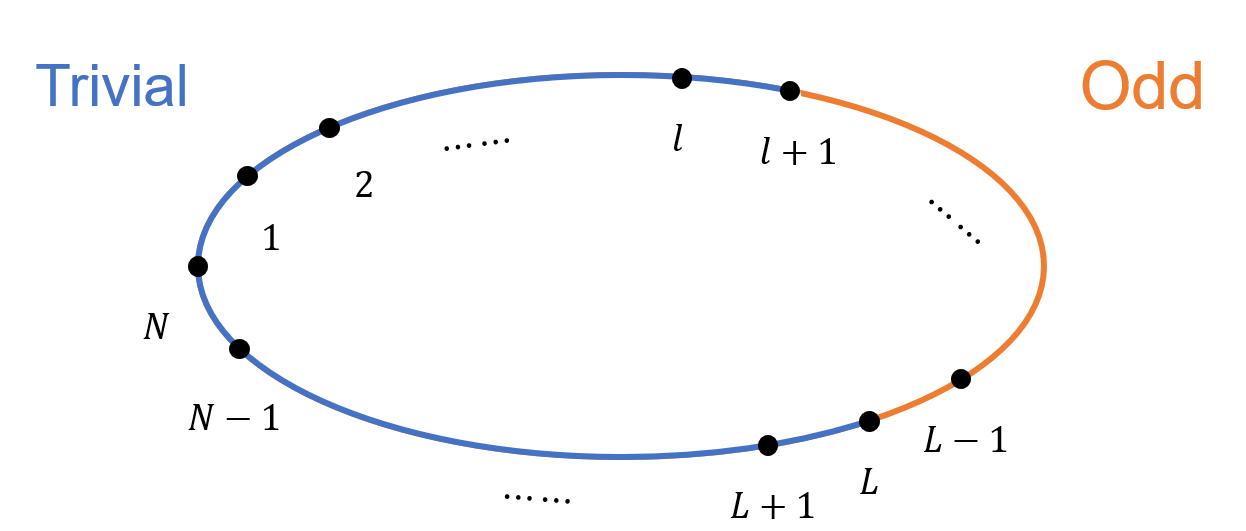}
         \caption{$H_{\rm trivial|odd}$ is defined on a spin chain of size $N$. It contains terms of a trivial SPT Hamiltonian from sites  $L+1,\ldots, N-1, N, 1\ldots l$, and an odd SPT Hamiltonian from sites $l+1 \ldots L$. Thus the model has two interfaces, one at $l$, and the other at $L$.}
        \label{fig:Floquet}
\end{figure}

\begin{figure*}
    \includegraphics[width=0.35\textwidth]{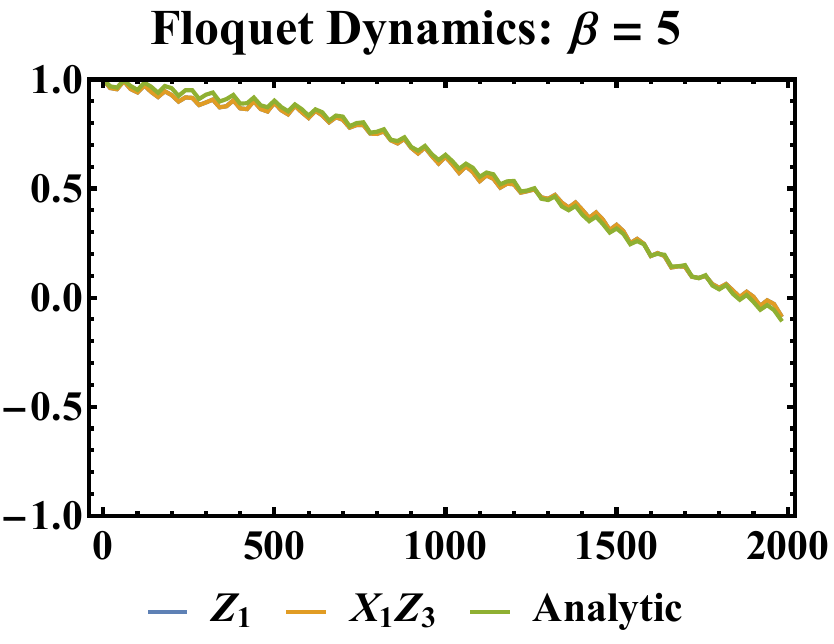}
    \includegraphics[width=0.35\textwidth]{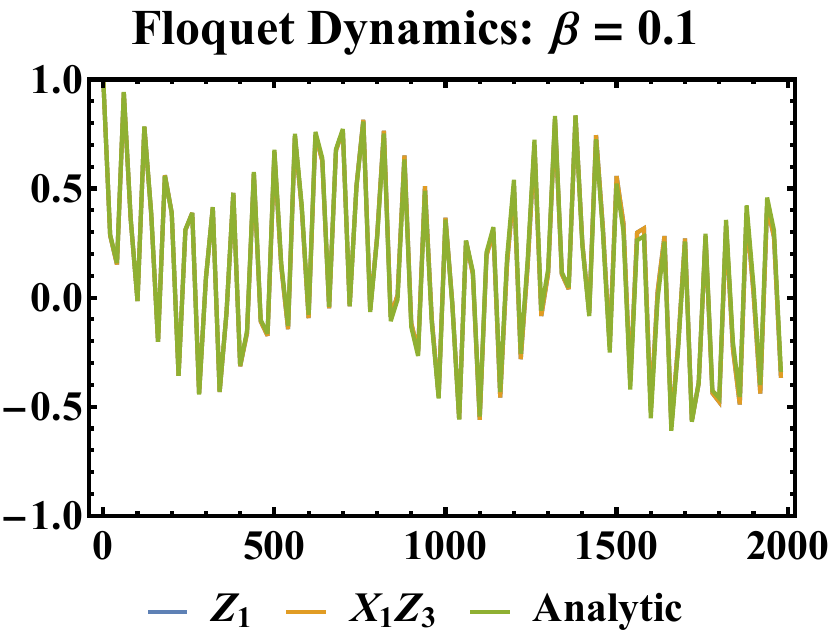}
    \includegraphics[width=0.35\textwidth]{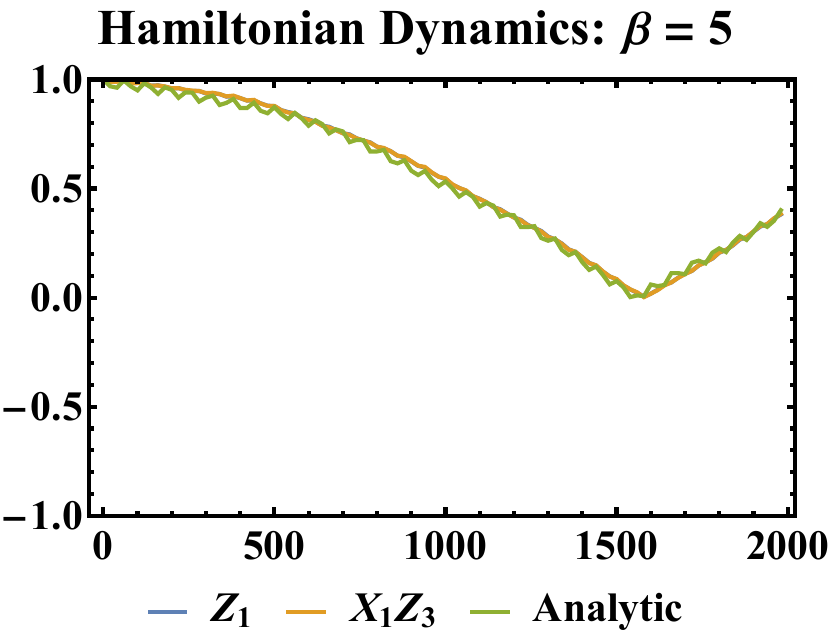}
    \includegraphics[width=0.35\textwidth]{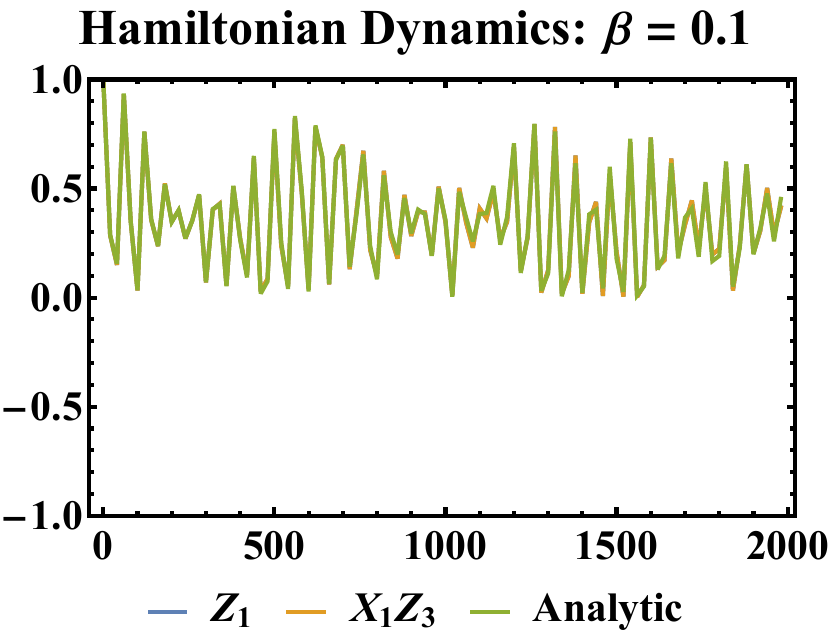}
    \caption{The absolute values of the autocorrelation functions of $X_1 Z_3$ and $Z_1$ operators for inverse temperature $\beta=0.1,5\ $ for Floquet \eqref{eq:AF} and  Hamiltonian  \eqref{eq:AH} dynamics. $J=1, B=0.1, L=12$ for all panels. The blue lines for $Z_1$ lie exactly under the orange lines for $X_1 Z_3$. The analytical results \eqref{eq:Aanal} match the numerical simulations.}
    \label{fig:num-analy}
\end{figure*}

In particular, we will now study the dynamics of edge modes when the system has Rep($D_8$) symmetry and has an interface that separates two regions with distinct ground-state SPTs. Let us consider a Floquet  unitary $F$ and a Hamiltonian $H$ where 
\beq
    F&=e^{-i\frac{J}{2}H_{\rm trivial|odd}}e^{-i\frac{B}{2}H_{\rm trivial}},\\ H&= \frac{J}{2}H_{\rm trivial|odd} + \frac{B}{2} H_{\rm trivial}. \label{eq:H-int}
\eeq
$H_{\rm trivial}$ is a trivial SPT Hamiltonian from sites $1\ldots N$. $H_{\rm trivial|odd}$ is a Hamiltonian that contains terms of a  trivial SPT Hamiltonian from sites $L+1,\ldots, N-1, N, 1\ldots l$, and an odd SPT Hamiltonian from sites $l+1 \ldots L$ (see Fig.~\ref{fig:Floquet}). The explicit forms of the Hamiltonians are
\begin{subequations}
\begin{align}
&H_{\rm trivial} = -\frac{1}{2}\sum_{j=1}^{N} X_{j},\\
& H_{\rm trivial|odd} = -\frac{1}{2}\sum_{j=L+1}^{l} X_{j}+\frac{1}{2}\sum_{n=1+l/2}^{L/2} X_{2n}\nonumber \\
    & +\frac{1}{2}\sum_{n=1+l/2}^{-2+L/2} Z_{2n-1}X_{2n+1}Z_{2n+3}\biggl(\frac{1+X_{2n}X_{2n+2}}{2}\biggr).
\end{align}
\end{subequations}

$H$ is the high-frequency ($B,J \ll 1$) Floquet Hamiltonian of $F$. $F$ can be  rewritten as 
\begin{align}
    F
    &=\biggl[e^{i\frac{B+J}{2}\sum_{j=L+1}^l X_j}e^{i\frac{B-J}{2}\sum_{n=1+l/2}^{L/2} X_{2n}}\biggl]\nonumber\\
    &\times \biggl[e^{-\frac{i}{2}\sum_{n=1+l/2}^{-2+L/2}\hat{J}_{2n} Z_{2n-1}X_{2n+1}Z_{2n+3}}e^{\frac{iB}{2}\sum_{n=l/2}^{-1+L/2} X_{2n+1}}\biggl],
\end{align}
where $\hat{J}_{2n}\equiv J(1+X_{2n}X_{2n+2})/2$. On applying a Jordan-Wigner transformation on the odd sites from $l+1$ to $L-1$,
\beq
    \gamma_{2k-1}=\biggl(\prod_{n=0}^{k-1}X_{2n+l+1}\biggr) Z_{2k+l+1};
    \gamma_{2k}= i X_{2k+l+1}\gamma_{2k-1}
    %=\prod_{n=0}^{k-1}X_{2n+l+1} \cdot Y_{2k+l+1},
\eeq
where $k=0,1,\cdots, (L-l)/2-1$,  $F$ becomes a fermion bi-linear in each symmetry sector labeled by $\hat{J}_{2n}$. Thus, one can explicitly solve for the eigenmodes. We first give the results for when all $X_{2n}=\pm 1$ so that $\hat{J}_{2n}=J$. In this case there are two zero modes on a single interface. These for $L\rightarrow \infty$ are (see Appendix \ref{appD})
\beq
    \Psi&\propto \sum_{n\geq 0}\left(-\frac{\tan \frac{B}{2}}{\tan \frac{J}{2}}\right)^n\left(\gamma_{4n-1}-\tan \frac{B}{2}\gamma_{4n}\right),\\
    \Psi'&\propto \sum_{n\geq 0}\left(-\frac{\tan \frac{B}{2}}{\tan \frac{J}{2}}\right)^n\left(\gamma_{4n+1}-\tan \frac{B}{2}\gamma_{4n+2}\right).
\eeq
When $B<J$, these two zero modes are localized near site $l+1$, and correspond to $Z_{l+1}$ and $X_{l+1} Z_{l+3}$ when $B=0$.  One may detect the interface mode by studying the autocorrelation function of the operator that has an overlap with the zero modes. Taking $l=0$, the natural choices are ${\mathcal O}=Z_1, X_1 Z_3$. We compute the following autocorrelation for Floquet (${\mathcal A}_F$), and Hamiltonian (${\mathcal A}_H$) dynamics,
\begin{subequations}
\begin{align}
 {\mathcal A}_F(n)&=\frac{1}{\Tr\left[e^{-\beta H'}\right]}\Tr\biggl[e^{-\beta H'}(F^{\dagger})^n {\mathcal O}F^n {\mathcal O}\biggr]
,\label{eq:AF}\\
{\mathcal A}_H(n)&=\frac{1}{\Tr\left[e^{-\beta H}\right]}\Tr\biggl[e^{-\beta H}e^{i H n}{\mathcal O} e^{-i H n}{\mathcal O}\biggr],\label{eq:AH}
\end{align}
\end{subequations}
where $H'=(J-B)\sum_{i} X_{2i}/2$. While temperature is not a natural concept for the Floquet problem, we introduce an effective-temperature via the weight $e^{-\beta H'}$ within the trace. This naturally highlights the role played by fluctuations in $\hat{J}_{2n}$ on the edge mode dynamics.

The dynamics involves summing over sectors $X_{2n}=\pm 1$ of the Hilbert space, where within each sector one has an effective cluster model, mappable to free fermions. The edge mode dynamics is dominated by sectors corresponding to cluster chains of different lengths, where a chain of length $k$ corresponds to $X_2=X_4=\cdots =X_{2k-2}\neq X_{2k}$. On a finite chain of length $k$, a zero mode of a homogeneous cluster chain is only approximate. Physically, there are localized zero modes on each boundary. Because of a nonzero overlap of order $(\frac{\tan B/2}{\tan J/2})^k$, the two zero modes interact, and the spectral degeneracy is lifted. 
Since the Hamiltonian dynamics is equivalent to the high frequency limit $B,J\ll 1 $, the overlap between the two zero modes of the Hamiltonian is $(B/J)^k$. 
Thus, the autocorrelation function can be approximated as
\begin{align}
&    {\mathcal A}_{F,H}(n) = \sum_{k=1\ldots L/2}p_k\cos\biggl[ \omega^{k-1}_{F,H} Bt\biggr],\nonumber\\
& \omega_{F} = \frac{\tan B/2}{\tan J/2},\omega_{H} = \frac{B}{J},\label{eq:Aanal}
\end{align}
with $p_k$ being the probability of having an effective chain length of $k$.
$\omega_{F,H}$ is related to the overlap between the zero modes. When $B\ll J$ this overlap is small, and the autocorrelation of the operator ${\mathcal O}$ evolves slowly.

 One can estimate $p_k$ as follows. 
 Each configuration $\{X_{2i}\}$ has an amplitude $e^{\beta(J- B) \sum_n s_{2n}}$, where $s_{2n}\equiv (1-X_{2n})/2$. Then  $p\equiv \frac{e^{\beta (J-B)}}{1+e^{\beta (J-B)}}$ is the amplitude for the spin on an even site to be in state $X=-1$. Thus, the 
 probability of a single site cluster chain is $p_1=2p(1-p)$, a $k$-site cluster chain is $p_k = p^{k}(1-p)+(1-p)^{k}p$ and that of a $L/2$ site cluster chain is $p^{L/2} + (1-p)^{L/2}$. 
In Fig.~\ref{fig:num-analy} we find that this approximate analytical result agrees well with the numerical simulation for $L=12$. 

Note that $\Psi,\Psi'$ anticommute with  the $\mathbb{Z}_2^2$ symmetry, while their product $\Psi_0 =\Psi \Psi'$ commutes with it. Thus, in the limit of $X_{2n}=\pm 1$
it is $\Psi_0$ which is a symmetry, and hence an exact zero mode. At non-zero temperature $\hat{J}_{2n}$ fluctuates, yet as shown above, there is still an approximate zero mode with slow dynamics at low effective temperatures. 

A hallmark of Floquet SPT is the ability to have edge modes with qualitatively different dynamics \cite{von2016phase,potter2016classification,potirniche2017floquet,bomantara2018quantum,Zhang2022Digital}. 
One natural candidate is the generalization of the zero mode to the $\pi$ mode, i.e, stable edge modes with period doubled dynamics. To construct such an edge mode for the Rep($D_8$) SPT, we first note that $KT$ while being non-invertible because it projects out odd $\mathbb{Z}_2^2$  symmetric sectors, can be made invertible if one studies the dynamics within the $\mathbb{Z}_2^2$ symmetric sector. Within this sector, the modified unitary $U=KT F$ can host a $\pi$ mode in the $X_{2n}=-1$ sector because $\Psi_0$ and $KT$ anti-commute. To see this, consider the simple limit of $B=0$, where $\Psi_0=Y_1 Z_3 $. Since, $KT Y_1 Z_3 = Y_1 X_2 Z_3 KT= -Y_1 Z_3 KT$ if $X_2=-1$, $\Psi_0$ anti-commutes with $KT$. The $\pi$ mode persists even with fluctuations in $X_{2n}$. This is because when the dynamics is weighted by $e^{-\beta H'}$, $X_{2n}=-1$ is weighted more than $X_{2n}=+1$ (see Appendix \ref{appD}). 

While zero and period-doubled $\pi$ modes  appear in many $\mathbb{Z}_2$ symmetric Floquet models such as Majorana chains and transverse field Ising models, we now discuss the difference between those modes and the ones presented here. In the former models the $0,\pi$ modes are charged under the invertible symmetry. In contrast, in Floquet problem studied here, the edge mode $\Psi_0=\Psi \Psi'$ is charged under the non-invertible $KT$ symmetry,  while commuting with the invertible $\mathbb{Z}_2\times \mathbb{Z}_2$ symmetry, in the low-energy sector. Hence the edge modes are protected by the non-invertible symmetry.

\section{ Conclusions} \label{concl}
Most studies of non-invertible symmetries focus on the ground state. Here we show that these symmetries have ramifications far out of equilibrium. They give rise to spectral degeneracies, co-existing trivial and non-trivial string order-parameters in excited states, and edge modes that oscillate slowly about zero and period doubled frequencies. While zero and period doubled edge modes arise for invertible Floquet SPTs, the ones constructed here differ from
the standard ones by being symmetric under the invertible symmetry, while being charged under the non-invertible symmetry.

Non-equilibrium phases that are stable to heating, require localization, and for small system sizes studied here, one cannot  differentiate between many-body localization, and pre-thermal physics \cite{abanin2017effective,yin2023prethermalization,long2023phenomenology}. In the thermodynamic limit, at the very least, we expect the stable dynamics presented here, to manifest for long times. Constructing string order-parameters for non-invertible Floquet SPTs is left for future work.

In Floquet models with ordinary invertible symmetries, localized edge modes with time crystalline behavior (period-doubled for the $\mathbb{Z}_2$ example) arise when one modifies a unitary $F \rightarrow U F$, where $U$ is a symmetry operator. Then if $F$ hosts a zero mode that is charged under $U$, $UF$ can host boundary time-crystals, with $F$ and $UF$ corresponding to different Floquet phases. $U$ can also be interpreted as a Thouless pump when studying the gapped ground states of related Hamiltonians. 

Recently Thouless pumps in SPT phases with non-invertible symmetries were classified~\cite{li2025classification}. As in the case of invertible symmetries, here too one can have two different Floquet unitaries that differ by a Thouless pump unitary. Exactly solvable non-fixed point Floquet models with these properties were constructed in Ref.~\cite{li2025classification}.

In contrast, in this paper we consider two different Floquet drives, $F$ and $KT F$, which differ by the non-invertible $KT$ symmetry. These also have the same effect in that if $F$ has zero modes, $KT F$ has $\pi$ modes. We expect that such one dimensional drives built out of unitaries and non-invertible symmetries can emerge naturally at the boundary of two dimensional topologically ordered systems. 

The Rep($D_8$) non-invertible SPT phases have Pauli stabilizer realizations, allowing us to generate all the excited states in the spectra, from the SPT ground state, using Pauli operators.  This approach can be generalized to non-invertible symmetries that are representation categories of the class-2 nilpotent group, whose SPT phases allow Pauli stabilizer realizations~\cite{li2024non}.
Generalizing this study to SPT phases for other non-invertible symmetries, currently known only in the form of commuting projectors~\cite{inamura2022lattice,meng2024non,bhardwaj2024lattice}, is an important open question.

{\sl Acknowledgments:} The authors thank Yifan Wang for helpful discussions. This work was supported by the U.S. National Science Foundation under Grant No. NSF DMR-2316598.

\appendix
\begin{widetext}
\section{Sequential circuit construction of the $KT$ duality}\label{appA}

In the sequential circuit construction, the $KT$ duality operator can be expressed as
\beq
    KT = T D_e D_o,
    \label{eq:sequential-circuit}
\eeq
where $T$ is the lattice translation by one site, and
\begin{subequations}
\begin{align}
    D_e=&e^{\frac{2\pi i L}{16}}\biggl(\frac{1+\eta_e}{ \sqrt{2}}\biggr)\frac{1-i Z_{1}X_L Z_{L-1}}{\sqrt{2}}\cdots \frac{1-iZ_4 Z_2}{\sqrt{2}}\frac{1-iZ_{3}X_{2}Z_{1}}{\sqrt{2}},\\
    D_o=&e^{\frac{2\pi i L}{16}}\biggl(\frac{1+\eta_o}{\sqrt{2}}\biggr)\frac{1-i Z_{L}X_{L-1} Z_{L-2}}{\sqrt{2}}\cdots \frac{1-iZ_3 Z_1}{\sqrt{2}}\frac{1-iZ_{2}X_{1}Z_{L}}{\sqrt{2}}.
\end{align}
The above sequential circuit can be obtained from the standard one that implements the Kramers-Wannier \cite{shao2023s} duality on conjugation  by Control-Z gates.

\end{subequations}

\section{Establishing ground state equivalence}
\label{appB}
We define a family of Hamiltonians built out of Rep($D_8$) symmetric operators. The simplest among them are 
\beq
    H_1^{(\pm, \pm)}=\sum_{i} \biggl[\pm X_{2i} \pm X_{2i+1}\biggr].
\eeq
Above we have used the convention that the first (second) signs in the superscript are tied to operators centered on the even (odd) sublattices. The next simplest sets of local commuting and symmetric operators give rise to the Hamiltonians,
\begin{align}
&    H_{2}^{(\pm, \pm)}=\pm \sum_{i}X_{2i}
    \pm \sum_{i}Z_{2i-1}X_{2i+1}Z_{2i+3}\biggl(\frac{1+X_{2i}X_{2i+2}}{2}\biggr),\label{H2pm}\\
&    H_{3}^{(\pm,\pm)}=\pm\sum_{i}Z_{2i-2}X_{2i}Z_{2i+2}\biggl(\frac{1+X_{2i-1}X_{2i+1}}{2}\biggr)\pm\sum_{i}X_{2i+1}.\label{H3pm}
\end{align}
The trivial SPT and the two non-trivial SPTs that are dubbed the odd, even SPTs are ground states of the following Hamiltonians
\beq
    H_{\rm{trivial}}\equiv H_1^{(-,-)}=-\sum_i X_i,\quad H_{\rm{odd},{\rm even}}=H_{2,3}^{(+,+)}.
    \label{eq:three SPTs}
\eeq
Of course, all the $H_1^{(\pm,\pm)}$ are trivial SPT Hamiltonians. 
Here we show a somewhat counterintuitive result that the signs of the terms in the Hamiltonians $H_{2,3}$ are crucial for specifying the phase.

The criterion for two Hamiltonians to be in the same SPT phase is that the ground states of two Hamiltonians can be related by a shallow depth circuit of locally symmetric unitary gates. For example, the following two Hamiltonians 
$    H_{\rm{trivial}}=-\sum_i X_i$ and $H_{\rm{cluster}}=-\sum_i Z_{i-1}X_i Z_{i+1}$
are in different SPT phases protected by the $\mathbb{Z}_2\times \mathbb{Z}_2$ symmetry generated by $\eta_e$ and $\eta_o$. However, if we only focus on the symmetry generated by $\eta_e \cdot \eta_o$, the two Hamiltonians belong to the same $\mathbb{Z}_2$ SPT phase, since the following circuit
\begin{align}
U=\prod_i \sqrt{Z_{2i-1}Z_{2i}}\left(\sqrt{Z_{2i}Z_{2i+1}}\right)^{\dagger}, 
\end{align}
maps between their ground states.

Let us consider the circuit 
\begin{align}
V_{o}=\prod_i X_{8i+1}X_{8i+3},
\end{align}
with Rep($D_8$) symmetric gates. This circuit maps the ground state of $H_{2}^{(+/-,-)}$ to the ground state of $H_{2}^{(+/-,+)}$, i.e, the mapping is between the two Hamiltonians that differ in the sign before the second term in \eqref{H2pm}.  This can be seen from the fact that $V_o$ flips the sign of only one of the $Z$ operators in the term $Z_{2i-1} X_{2i+1}Z_{2i+3}$. 
Since the ground state of $H_{2}^{(+,+)}$ is the non-trivial odd SPT, this implies that the ground state of $H_2^{(+,-)}$ is also a non-trivial SPT.  We will show below that the ground state of $H_2^{(-,+)}$ is a trivial SPT, and from the above argument, so is the ground state of $H_2^{(-,-)}$. 

Similarly, 
\begin{align}
V_{e}=\prod_i X_{8i}X_{8i+2},
\end{align}
maps the ground state of $H_{3}^{(-,+/-)}$ to the ground state of $H_{3}^{(+,+/-)}$, where the two Hamiltonians differ in the sign before the first term in \eqref{H3pm}. Since the ground state of $H_{3}^{(+,+)}$ is the non-trivial even SPT, this implies that the ground state of $H_3^{(-,+)}$ is a non-trivial SPT.  We will show below that the ground state of $H_3^{(+,-)}$ is a trivial SPT, and from the above argument, so is the ground state of $H_3^{(-,-)}$.

Let us consider the circuit 
\begin{align}
U_{o}=\prod_i \sqrt{u_{4i}}\left(\sqrt{u_{4i+2}}\right)^{\dagger}, u_{j}\equiv \frac{(1+X_{j})Z_{j-1}Z_{j+1}}{2}+\frac{(1-X_{j})}{2}.
\end{align}
The above circuit, maps the ground state of $H_{\rm trivial}=H_{1}^{(-,+/-)}$ to the ground state of $H_{2}^{(-,+/-)}$.  To see this, we note that when a gate $u_{j}$ acts on the ground state, since $X_{j}=1$ for any even site $j$, $u_{j}$ is effectively just $\sqrt{Z_{j-1}Z_{j+1}}$, reducing $U_o$ to $U$ on the odd sublattice, and mapping the trivial product state to the cluster state.  Similarly, the circuit
$   U_{e}=\prod_i \sqrt{u_{4i-1}}\left(\sqrt{u_{4i+1}}\right)^{\dagger}$
maps between the ground states of $H_1^{(+/-,-)}$ and $H_3^{(+/-,-)}$. Thus the ground states of $H_2^{(-,+/-)},H_3^{(+/-,-)}$ can be trivial SPTs.

Thus to summarize, $H_{\rm odd}=H_2^{(+,+)}, H_{\rm even}=H_3^{(+,+)}$ have ground states that are non-trivial SPTs. In contrast, $H_2^{(-,\pm)}, H_3^{(\pm,-)}$ have trivial SPT ground states.  

\section{MPO derivations} \label{appC}
The matrix product operator (MPO) construction of $KT$ is as follows
\beq
    KT= \sum_{i_1,j_1,\cdots,i_L,j_L}\Tr(A^{i_i}_{j_1} A^{i_2}_{j_2} \cdots A^{i_L}_{j_L})\ket{i_1,i_2,\cdots, i_L}\bra{j_1,j_2,\cdots,j_L},
\eeq
where the tensor $A$, for a set of physical indices, is the following bond dimension two matrix in virtual space
\begin{align*}
\begin{tikzpicture}[baseline=(current bounding box),scale=1.2]
        \draw[string] (0,1.3)node[below]{$0$} -- (0,1.7);
        \draw[string] (0,2.3) -- (0,2.7)node[above]{$0$};
        \filldraw[fill = white] (-.3,1.7) rectangle node {$A$} (.3,2.3);
        \draw[string] (0.7,2) --  (0.3,2);
        \draw[string] (-0.3,2) --  (-0.7,2);
\end{tikzpicture}=
\begin{tikzpicture}[baseline=(current bounding box),scale=1.2]
        \draw[string] (0,1.3)node[below]{$1$} -- (0,1.7);
        \draw[string] (0,2.3) -- (0,2.7)node[above]{$1$};
        \filldraw[fill = white] (-.3,1.7) rectangle node {$A$} (.3,2.3);
        \draw[string] (0.7,2) --  (0.3,2);
        \draw[string] (-0.3,2) --  (-0.7,2);
\end{tikzpicture}=\ket{+}\bra{0},\quad
\begin{tikzpicture}[baseline=(current bounding box),scale=1.2]
        \draw[string] (0,1.3)node[below]{$0$} -- (0,1.7);
        \draw[string] (0,2.3) -- (0,2.7)node[above]{$1$};
        \filldraw[fill = white] (-.3,1.7) rectangle node {$A$} (.3,2.3);
        \draw[string] (0.7,2) --  (0.3,2);
        \draw[string] (-0.3,2) --  (-0.7,2);
\end{tikzpicture}=
\begin{tikzpicture}[baseline=(current bounding box),scale=1.2]
        \draw[string] (0,1.3)node[below]{$1$} -- (0,1.7);
        \draw[string] (0,2.3) -- (0,2.7)node[above]{$0$};
        \filldraw[fill = white] (-.3,1.7) rectangle node {$A$} (.3,2.3);
        \draw[string] (0.7,2) --  (0.3,2);
        \draw[string] (-0.3,2) --  (-0.7,2);
\end{tikzpicture}=\ket{-}\bra{1}.
\end{align*}
It is straightforward to show that the tensor $A$ satisfies
\begin{align*}
\begin{tikzpicture}[baseline=(current bounding box),scale=1.2]
        \draw[string] (0,1.3) -- (0,1.7);
        \draw[string] (0,2.3) -- (0,2.7);
        \filldraw[fill = white] (-.3,1.7) rectangle node {$A$} (.3,2.3);
        \draw[string] (0.7,2) --  (0.3,2);
        \draw[string] (-0.3,2) --  (-0.7,2);
\end{tikzpicture}\ = \
\begin{tikzpicture}[baseline=(current bounding box),scale=1.2]
        \draw[string] (0,1.3) -- (0,1.7);
        \draw[string] (0,2.3) -- (0,2.7);
        \filldraw[fill = white] (-.3,1.7) rectangle node {$A$} (.3,2.3);
        \draw[string] (0.7,2) --  (0.3,2);
        \draw[string] (-0.3,2) --  (-0.7,2);
        \filldraw[fill = white] (-.2,2.7) rectangle node {$X$} (.2,3.1);
        \draw[string] (0,3.1) -- (0,3.5);
        \filldraw[fill = white] (-.2,0.9) rectangle node {$X$} (.2,1.3);
        \draw[string] (0,0.5) -- (0,0.9);
\end{tikzpicture}\ = \
\begin{tikzpicture}[baseline=(current bounding box),scale=1.2]
        \draw[string] (0,1.3) -- (0,1.7);
        \draw[string] (0,2.3) -- (0,2.7);
        \filldraw[fill = white] (-.3,1.7) rectangle node {$A$} (.3,2.3);
        \draw[string] (0.7,2) --  (0.3,2);
        \draw[string] (-0.3,2) --  (-0.7,2);
        \filldraw[fill = white] (-.2,2.7) rectangle node {$Z$} (.2,3.1);
        \draw[string] (0,3.1) -- (0,3.5);
        \filldraw[fill = white] (-.2,0.9) rectangle node {$Z$} (.2,1.3);
        \draw[string] (0,0.5) -- (0,0.9);
        \filldraw[fill = white] (0.7,1.8) rectangle node {$Z$} (1.1,2.2);
        \draw[string] (1.5,2) -- (1.1,2);
\end{tikzpicture}\ = \
\begin{tikzpicture}[baseline=(current bounding box),scale=1.2]
        \draw[string] (0,1.3) -- (0,1.7);
        \draw[string] (0,2.3) -- (0,2.7);
        \filldraw[fill = white] (-.3,1.7) rectangle node {$A$} (.3,2.3);
        \draw[string] (0.7,2) --  (0.3,2);
        \draw[string] (-0.3,2) --  (-0.7,2);
        \filldraw[fill = white] (-.2,2.7) rectangle node {$Z$} (.2,3.1);
        \draw[string] (0,3.1) -- (0,3.5);
        \filldraw[fill = white] (-.2,0.9) rectangle node {$Z$} (.2,1.3);
        \draw[string] (0,0.5) -- (0,0.9);
        \filldraw[fill = white] (-1.1,1.8) rectangle node {$X$} (-0.7,2.2);
        \draw[string] (-1.1,2) -- (-1.5,2);
\end{tikzpicture}\ = \
\begin{tikzpicture}[baseline=(current bounding box),scale=1.2]
        \draw[string] (0,1.3) -- (0,1.7);
        \draw[string] (0,2.3) -- (0,2.7);
        \filldraw[fill = white] (-.3,1.7) rectangle node {$A$} (.3,2.3);
        \draw[string] (0.7,2) --  (0.3,2);
        \draw[string] (-0.3,2) --  (-0.7,2);
        \filldraw[fill = white] (-.2,2.7) rectangle node {$X$} (.2,3.1);
        \draw[string] (0,3.1) -- (0,3.5);
        \filldraw[fill = white] (-1.1,1.8) rectangle node {$Z$} (-0.7,2.2);
        \draw[string] (-1.1,2) -- (-1.5,2);
        \filldraw[fill = white] (0.7,1.8) rectangle node {$X$} (1.1,2.2);
        \draw[string] (1.5,2) -- (1.1,2);
\end{tikzpicture},
\end{align*}
from which the following operator relations follow,
\beq
    KT \cdot X_i = X_i \cdot KT,\quad KT\cdot Z_{i-1}Z_{i+1} = Z_{i-1}X_{i}Z_{i+1}\cdot KT.
\eeq
The ground state of $H_{\rm trivial}$, i.e. $\ket{+}^{\otimes L} $ is invariant under the action of $KT$. To see this, we first write this state as a matrix product state (MPS)
\beq
    \ket{+}^{\otimes L}  = \sum_{i_1,\cdots,i_L}\Tr(B^{i_i}B^{i_2} \cdots B^{i_L})\ket{i_1,i_2,\cdots, i_L},
\eeq
where $B^0 = B^1 = \frac{1}{\sqrt{2}}$ is just a scalar (virtual bond dimension is zero). The state $KT\ket{+}^{\otimes L}$ is an MPS given by the following matrix,
\begin{align*}
\begin{tikzpicture}[baseline=(current bounding box),scale=1.2]
        \draw[string] (0,1.3) -- (0,1.7);
        \draw[string] (0,2.3) -- (0,2.7);
        \filldraw[fill = white] (-.3,1.7) rectangle node {$A$} (.3,2.3);
        \draw[string] (0.7,2) --  (0.3,2);
        \draw[string] (-0.3,2) --  (-0.7,2);
        \filldraw[fill = white] (-.3,0.7) rectangle node {$B$} (.3,1.3);
\end{tikzpicture}=
\begin{tikzpicture}[baseline=(current bounding box),scale=1.2]
        \draw[string] (0,2.3) -- (0,2.7);
        \filldraw[fill = white] (-.3,1.7) rectangle node {$B$} (.3,2.3);
        \draw[string] (0.7,1) --  (0.3,1);
        \draw[string] (-0.3,1) --  (-0.7,1);
        \filldraw[fill = white] (-.3,0.7) rectangle node {$H$} (.3,1.3);
\end{tikzpicture}
\end{align*}
where $H\equiv \ket{+}\bra{0}+\ket{-}\bra{1}$ is the Hadamard gate. Hence, $KT \ket{+}^{\otimes L} = 2\ket{+}^{\otimes L}$.

Now we study the ground state $\ket{\rm odd}$ of $H_{\rm odd}$. The state $\ket{\rm odd}$ is given by the MPS
\begin{align*}
\begin{tikzpicture}[baseline=(current bounding box),scale=1.2]
        \node[text width=3cm] at (-5,0){$\cdots$};
        \node[text width=3cm] at (2.1,0){$\cdots$};
        \draw[string] (-5,1.3) -- (-5,1.7);
        \filldraw[fill = white] (-5.3,0.7) rectangle node {$C$} (-4.7,1.3);
% ----------------
        \draw[string] (-4,.3) -- (-4,1.7);
        \filldraw[fill = white] (-4.3,-.3) rectangle node {$D$} (-3.7,.3);
        \draw[string] (-4.3,0) --  (-5.7,0);
%-------------------    
        \draw[string] (-3,1.3) -- (-3,1.7);
        \filldraw[fill = white] (-3.3,0.7) rectangle node {$C$} (-2.7,1.3);
% ----------------
        \draw[string] (-2,.3) -- (-2,1.7);
        \filldraw[fill = white] (-2.3,-.3) rectangle node {$D$} (-1.7,.3);
        \draw[string] (-2.3,0) --  (-3.7,0);
%-------------------
        \draw[string] (-1,1.3) -- (-1,1.7);
        \filldraw[fill = white] (-1.3,0.7) rectangle node {$C$} (-0.7,1.3);
% ----------------
        \draw[string] (0,.3) -- (0,1.7);
        \filldraw[fill = white] (-.3,-.3) rectangle node {$D$} (.3,.3);
        \draw[string] (0.7,0) --  (0.3,0);
        \draw[string] (-0.3,0) --  (-1.7,0);
\end{tikzpicture}
\end{align*}
where $C^0 = -C^1 = \frac{1}{\sqrt{2}}$, $D^0=\ket{+}\bra{0}$, and $D^1=-\ket{-}\bra{1}$. After the action of $KT$, the state $KT\ket{\rm odd}$ can be written as follows,
\begin{align*}
&\begin{tikzpicture}[baseline=(current bounding box),scale=1.2]
        \node[text width=3cm] at (-5,2){$\cdots$};
        \node[text width=3cm] at (-5,0){$\cdots$};
        \node[text width=3cm] at (2.1,2){$\cdots$};
        \node[text width=3cm] at (2.1,0){$\cdots$};
        \draw[string] (-5,1.3) -- (-5,1.7);
        \draw[string] (-5,2.3) -- (-5,2.7);
        \filldraw[fill = white] (-5.3,1.7) rectangle node {$A$} (-4.7,2.3);
        \draw[string] (-4.3,2) --  (-4.7,2);
        \draw[string] (-5.3,2) --  (-5.7,2);
        \filldraw[fill = white] (-5.3,0.7) rectangle node {$C$} (-4.7,1.3);
% ----------------
        \draw[string] (-4,.3) -- (-4,1.7);
        \draw[string] (-4,2.3) -- (-4,2.7);
        \filldraw[fill = white] (-4.3,1.7) rectangle node {$A$} (-3.7,2.3);
        \draw[string] (-3.3,2) --  (-3.7,2);
        \draw[string] (-4.3,2) --  (-4.7,2);
        \filldraw[fill = white] (-4.3,-.3) rectangle node {$D$} (-3.7,.3);
        \draw[string] (-4.3,0) --  (-5.7,0);
%-------------------    
        \draw[string] (-3,1.3) -- (-3,1.7);
        \draw[string] (-3,2.3) -- (-3,2.7);
        \filldraw[fill = white] (-3.3,1.7) rectangle node {$A$} (-2.7,2.3);
        \draw[string] (-2.3,2) --  (-2.7,2);
        \draw[string] (-3.3,2) --  (-3.7,2);
        \filldraw[fill = white] (-3.3,0.7) rectangle node {$C$} (-2.7,1.3);
% ----------------
        \draw[string] (-2,.3) -- (-2,1.7);
        \draw[string] (-2,2.3) -- (-2,2.7);
        \filldraw[fill = white] (-2.3,1.7) rectangle node {$A$} (-1.7,2.3);
        \draw[string] (-1.3,2) --  (-1.7,2);
        \draw[string] (-2.3,2) --  (-2.7,2);
        \filldraw[fill = white] (-2.3,-.3) rectangle node {$D$} (-1.7,.3);
        \draw[string] (-2.3,0) --  (-3.7,0);
%-------------------
        \draw[string] (-1,1.3) -- (-1,1.7);
        \draw[string] (-1,2.3) -- (-1,2.7);
        \filldraw[fill = white] (-1.3,1.7) rectangle node {$A$} (-0.7,2.3);
        \draw[string] (-0.3,2) --  (-0.7,2);
        \draw[string] (-1.3,2) --  (-1.7,2);
        \filldraw[fill = white] (-1.3,0.7) rectangle node {$C$} (-0.7,1.3);
% ----------------
        \draw[string] (0,.3) -- (0,1.7);
        \draw[string] (0,2.3) -- (0,2.7);
        \filldraw[fill = white] (-.3,1.7) rectangle node {$A$} (.3,2.3);
        \draw[string] (0.7,2) --  (0.3,2);
        \draw[string] (-0.3,2) --  (-0.7,2);
        \filldraw[fill = white] (-.3,-.3) rectangle node {$D$} (.3,.3);
        \draw[string] (0.7,0) --  (0.3,0);
        \draw[string] (-0.3,0) --  (-1.7,0);
\end{tikzpicture}\\ 
\\
=\ & 
\begin{tikzpicture}[baseline=(current bounding box),scale=1.2]
        \node[text width=3cm] at (-5,1){$\cdots$};
        \node[text width=3cm] at (-5,0){$\cdots$};
        \node[text width=3cm] at (2.1,1){$\cdots$};
        \node[text width=3cm] at (2.1,0){$\cdots$};
        \draw[string] (-5,2.3) -- (-5,2.7);
        \filldraw[fill = white] (-5.3,1.7) rectangle node {$C$} (-4.7,2.3);
        \draw[string] (-4.3,1) --  (-4.7,1);
        \draw[string] (-5.3,1) --  (-5.7,1);
        \filldraw[fill = white] (-5.3,0.7) rectangle node {$X$} (-4.7,1.3);
% ----------------
        \draw[string] (-4,1.3) -- (-4,2.7);
        \draw[string] (-4,.3) -- (-4,0.7);
        \filldraw[fill = white] (-4.3,0.7) rectangle node {$\Lambda$} (-3.7,1.3);
        \draw[string] (-3.3,1) --  (-3.7,1);
        \draw[string] (-4.3,1) --  (-4.7,1);
        \filldraw[fill = white] (-4.3,-.3) rectangle node {$D$} (-3.7,.3);
        \draw[string] (-4.3,0) --  (-5.7,0);
%-------------------    
        \draw[string] (-3,2.3) -- (-3,2.7);
        \filldraw[fill = white] (-3.3,1.7) rectangle node {$C$} (-2.7,2.3);
        \draw[string] (-2.3,1) --  (-2.7,1);
        \draw[string] (-3.3,1) --  (-3.7,1);
        \filldraw[fill = white] (-3.3,0.7) rectangle node {$X$} (-2.7,1.3);
% ----------------
        \draw[string] (-2,1.3) -- (-2,2.7);
        \draw[string] (-2,.3) -- (-2,0.7);
        \filldraw[fill = white] (-2.3,.7) rectangle node {$\Lambda$} (-1.7,1.3);
        \draw[string] (-1.3,1) --  (-1.7,1);
        \draw[string] (-2.3,1) --  (-2.7,1);
        \filldraw[fill = white] (-2.3,-.3) rectangle node {$D$} (-1.7,.3);
        \draw[string] (-2.3,0) --  (-3.7,0);
%-------------------
        \draw[string] (-1,2.3) -- (-1,2.7);
        \filldraw[fill = white] (-1.3,1.7) rectangle node {$C$} (-0.7,2.3);
        \draw[string] (-0.3,1) --  (-0.7,1);
        \draw[string] (-1.3,1) --  (-1.7,1);
        \filldraw[fill = white] (-1.3,0.7) rectangle node {$X$} (-0.7,1.3);
% ----------------
        \draw[string] (0,.3) -- (0,.7);
        \draw[string] (0,1.3) -- (0,2.7);
        \filldraw[fill = white] (-.3,.7) rectangle node {$\Lambda$} (.3,1.3);
        \draw[string] (0.7,1) --  (0.3,1);
        \draw[string] (-0.3,1) --  (-0.7,1);
        \filldraw[fill = white] (-.3,-.3) rectangle node {$D$} (.3,.3);
        \draw[string] (0.7,0) --  (0.3,0);
        \draw[string] (-0.3,0) --  (-1.7,0);
\end{tikzpicture}
\end{align*}
where we used the relation,
\begin{align*}
\begin{tikzpicture}[baseline=(current bounding box),scale=1.2]
        \draw[string] (0,1.3) -- (0,1.7);
        \draw[string] (0,2.3) -- (0,2.7);
        \filldraw[fill = white] (-.3,1.7) rectangle node {$A$} (.3,2.3);
        \draw[string] (0.7,2) --  (0.3,2);
        \draw[string] (-0.3,2) --  (-0.7,2);
        \filldraw[fill = white] (-.3,0.7) rectangle node {$C$} (.3,1.3);
\end{tikzpicture}\ =\
\begin{tikzpicture}[baseline=(current bounding box),scale=1.2]
        \draw[string] (0,2.3) -- (0,2.7);
        \filldraw[fill = white] (-.3,1.7) rectangle node {$C$} (.3,2.3);
        \draw[string] (0.7,1) --  (0.3,1);
        \draw[string] (-0.3,1) --  (-0.7,1);
        \filldraw[fill = white] (-.3,0.7) rectangle node {$X$} (.3,1.3);
        \filldraw[fill = white] (0.7,.7) rectangle node {$H$} (1.3,1.3);
        \draw[string] (1.7,1) --  (1.3,1);
\end{tikzpicture}
\end{align*}
and the tensor $\Lambda$ is defined as
\begin{align*}
\begin{tikzpicture}[baseline=(current bounding box),scale=1.2]
        \draw[string] (0,1.3) -- (0,1.7);
        \draw[string] (0,2.3) -- (0,2.7);
        \filldraw[fill = white] (-.3,1.7) rectangle node {$\Lambda$} (.3,2.3);
        \draw[string] (0.7,2) --  (0.3,2);
        \draw[string] (-0.3,2) --  (-0.7,2);
\end{tikzpicture}\ \equiv\
\begin{tikzpicture}[baseline=(current bounding box),scale=1.2]
        \draw[string] (0,1.3) -- (0,1.7);
        \draw[string] (0,2.3) -- (0,2.7);
        \filldraw[fill = white] (-.3,1.7) rectangle node {$A$} (.3,2.3);
        \draw[string] (0.7,2) --  (0.3,2);
        \draw[string] (-0.3,2) --  (-0.7,2);
        \filldraw[fill = white] (-1.3,1.7) rectangle node {$H$} (-0.7,2.3);
        \draw[string] (-1.3,2) --  (-1.7,2);
\end{tikzpicture}
\end{align*}
and satisfies $\Lambda^0_0=\Lambda^1_1 = \ket{0}\bra{0}$ and $\Lambda^0_1=\Lambda^1_0 = \ket{1}\bra{1}$. 
As a result, 
\beq
    KT\ket{\rm odd}=(\prod_i X_{4i+1} + \prod_i X_{4i+3})\ket{\rm odd}= 2\ket{\rm odd}.
\eeq
Now we discuss the string order parameters,
\beq
    (^{\text{trivial}}\mathcal{O}^{\text{string}}_{KT})_{i,j; I, J} &= \bra{i}A_{2I}\cdots A_{2J+1}\ket{j},\\
 (^{\text{Odd}}\mathcal{O}^{\text{string}}_{KT})_{i,j; I, J} &= Z_{2I-3}^{i}X_{2I-1}^{i}Z_{2I-1}^{1-i}\bra{i}A_{2I}\cdots A_{2J+1}\ket{j} Z_{2J+3}^{j}X_{2J+3}^{1-j}Z_{2J+5}^{1-j}.     
\eeq
They contain $\bra{i}A_{2I}\cdots A_{2J+1}\ket{j}$, which can be simplified in the same way as above,
\begin{align*}
&\begin{tikzpicture}[baseline=(current bounding box),scale=1.2]
        \node[text width=3cm] at (-4.8,2){$\bra{i}$};
        \node[text width=3cm] at (-5,0){$\cdots$};
        \node[text width=3cm] at (2,2){$\ket{j}$};
        \node[text width=3cm] at (2.1,0){$\cdots$};
        \draw[string] (-5,1.3) -- (-5,1.7);
        \draw[string] (-5,2.3) -- (-5,2.7);
        \filldraw[fill = white] (-5.3,1.7) rectangle node {$A$} (-4.7,2.3);
        \draw[string] (-4.3,2) --  (-4.7,2);
        \draw[string] (-5.3,2) --  (-5.7,2);
        \filldraw[fill = white] (-5.3,0.7) rectangle node {$C$} (-4.7,1.3);
% ----------------
        \draw[string] (-4,.3) -- (-4,1.7);
        \draw[string] (-4,2.3) -- (-4,2.7);
        \filldraw[fill = white] (-4.3,1.7) rectangle node {$A$} (-3.7,2.3);
        \draw[string] (-3.3,2) --  (-3.7,2);
        \draw[string] (-4.3,2) --  (-4.7,2);
        \filldraw[fill = white] (-4.3,-.3) rectangle node {$D$} (-3.7,.3);
        \draw[string] (-4.3,0) --  (-5.7,0);
        \draw[string] (-3.3,0) --  (-3.7,0);
%-------------------    
        \node[text width=3cm] at (-1.6,2){$\cdots\cdots$};
        \node[text width=3cm] at (-1.6,0){$\cdots\cdots$};
%-------------------
        \draw[string] (-1,1.3) -- (-1,1.7);
        \draw[string] (-1,2.3) -- (-1,2.7);
        \filldraw[fill = white] (-1.3,1.7) rectangle node {$A$} (-0.7,2.3);
        \draw[string] (-0.3,2) --  (-0.7,2);
        \draw[string] (-1.3,2) --  (-1.7,2);
        \filldraw[fill = white] (-1.3,0.7) rectangle node {$C$} (-0.7,1.3);
% ----------------
        \draw[string] (0,.3) -- (0,1.7);
        \draw[string] (0,2.3) -- (0,2.7);
        \filldraw[fill = white] (-.3,1.7) rectangle node {$A$} (.3,2.3);
        \draw[string] (0.7,2) --  (0.3,2);
        \draw[string] (-0.3,2) --  (-0.7,2);
        \filldraw[fill = white] (-.3,-.3) rectangle node {$D$} (.3,.3);
        \draw[string] (0.7,0) --  (0.3,0);
        \draw[string] (-0.3,0) --  (-1.7,0);
\end{tikzpicture}\\ \\
=\ &\begin{tikzpicture}[baseline=(current bounding box),scale=1.2]
        \node[text width=3cm] at (-5,0){$\cdots$};
        \node[text width=3cm] at (-4.8,1){$\bra{i}$};
        \node[text width=3cm] at (2,1){$\ket{j}$};
        \node[text width=3cm] at (2.1,0){$\cdots$};
        \draw[string] (-5,2.3) -- (-5,2.7);
        \filldraw[fill = white] (-5.3,1.7) rectangle node {$C$} (-4.7,2.3);
        \draw[string] (-4.3,1) --  (-4.7,1);
        \draw[string] (-5.3,1) --  (-5.7,1);
        \filldraw[fill = white] (-5.3,0.7) rectangle node {$X$} (-4.7,1.3);
% ----------------
        \draw[string] (-4,1.3) -- (-4,2.7);
        \draw[string] (-4,.3) -- (-4,0.7);
        \filldraw[fill = white] (-4.3,0.7) rectangle node {$\Lambda$} (-3.7,1.3);
        \draw[string] (-3.3,1) --  (-3.7,1);
        \draw[string] (-4.3,1) --  (-4.7,1);
        \filldraw[fill = white] (-4.3,-.3) rectangle node {$D$} (-3.7,.3);
        \draw[string] (-4.3,0) --  (-5.7,0);  
        \draw[string] (-3.3,0) --  (-3.7,0);
%-------------------    
        \node[text width=3cm] at (-1.6,1){$\cdots\cdots$};
        \node[text width=3cm] at (-1.6,0){$\cdots\cdots$};
%-------------------
        \draw[string] (-1,2.3) -- (-1,2.7);
        \filldraw[fill = white] (-1.3,1.7) rectangle node {$C$} (-0.7,2.3);
        \draw[string] (-0.3,1) --  (-0.7,1);
        \draw[string] (-1.3,1) --  (-1.7,1);
        \filldraw[fill = white] (-1.3,0.7) rectangle node {$X$} (-0.7,1.3);
% ----------------
        \draw[string] (0,.3) -- (0,.7);
        \draw[string] (0,1.3) -- (0,2.7);
        \filldraw[fill = white] (-.3,.7) rectangle node {$\Lambda$} (.3,1.3);
        \draw[string] (0.7,1) --  (0.3,1);
        \draw[string] (-0.3,1) --  (-0.7,1);
        \filldraw[fill = white] (-.3,-.3) rectangle node {$D$} (.3,.3);
        \draw[string] (0.7,0) --  (0.3,0);
        \draw[string] (-0.3,0) --  (-1.7,0);
\end{tikzpicture}
\end{align*}
From the last line of the above equation we can see that, when $J-I$ is odd, 
\beq
    \bra{i}A_{2I}\cdots A_{2J+1}\ket{j}\ket{\rm odd} = \begin{cases}
    \prod_{k=0,\cdots,\frac{J-I-1}{2}}  X_{2I+1+4k}^{1-i}X_{2I+3+4k}^{i}\ket{\rm odd}, & \text{if } i = j, \\
    0, & \text{if } i \neq j.
\end{cases}
\eeq
When $J-I$ is even, 
\beq
    \bra{i}A_{2I}\cdots A_{2J+1}\ket{j}\ket{\rm odd} = \begin{cases}
    X_{2I+1}\cdots X_{2J-3}X_{2J+1}\ket{\rm odd}, & \text{if } i = 0, j = 1, \\
    X_{2I+3}\cdots X_{2J-5}X_{2J-1}\ket{\rm odd}, & \text{if } i = 1, j = 0 \\
    0, & \text{if } i = j.
\end{cases}
\eeq
From the above results, we can show 
\beq
    \bra{\rm odd}(^{\text{trivial}}\mathcal{O}^{\text{string}}_{KT})_{i,j; I, J}\ket{\rm odd} = 0.
\eeq
If we write the expectation value of $(^{\text{odd}}\mathcal{O}^{\text{string}}_{KT})_{i,j; I, J}$ as a $2\times 2$ matrix with indices $i$ and $j$, then
\beq
    \bra{\rm odd}(^{\text{\rm odd}}\mathcal{O}^{\text{string}}_{KT})_{I, J}\ket{\rm odd} = \begin{cases}
    \begin{pmatrix}
        1 & 0 \\
        0 & 1
    \end{pmatrix}, & \text{if } $I-J$ \text{ is odd}, \\
    \\
    \begin{pmatrix}
        0 & 1 \\
        1 & 0
    \end{pmatrix}, & \text{if } $I-J$ \text{ is even}.
\end{cases}
\eeq

\section{Zero and $\pi$ edge modes}\label{appD}
Consider $F$ defined in the main text. We apply the Jordan-Wigner transformation on the odd sites from $l+1$ to $L-1$,
\beq
    \gamma_{2k-1}=\prod_{n=0}^{k-1}X_{2n+l+1} \cdot Z_{2k+l+1},\quad \gamma_{2k}=\prod_{n=0}^{k-1}X_{2n+l+1} \cdot Y_{2k+l+1},
\eeq
$k=0,1,\cdots, \frac{L-l}{2}-1$. The Floquet drive becomes
\beq
    F=\exp(i\sum_{n=1+l/2}^{L/2} \frac{B-J}{2}X_{2n})\exp(\sum_{k=0}^{\frac{L-l}{2}-3}\frac{\hat{J}_{2k+l+2}}{2} \gamma_{2k}\gamma_{2k+3})\exp(-\sum_{k=0}^{\frac{L-l}{2}-1} \frac{B}{2} \gamma_{2k-1}\gamma_{2k}),
\eeq
which is formed by fermions bi-linears. Therefore $F\gamma_j F^{\dagger}$ is a linear combination of the $\gamma$ operators.
Using $e^{A \gamma_a \gamma_b/2}\gamma_a e^{-A \gamma_a \gamma_b/2}=\cos(A) \gamma_a -\sin(A)\gamma_b $ we have 
\beq
    \gamma_{2k-1}&\longrightarrow c_{k-2}c'\gamma_{2k-1}+s_{k-2}c'\gamma_{2k-4}+c_k s'\gamma_{2k}-s_k s'\gamma_{2k+3},\\
    \gamma_{2k}&\longrightarrow c_k c'\gamma_{2k}-s_k c'\gamma_{2k+3}-c_{k-2}s'\gamma_{2k-1}-s_{k-2}s'\gamma_{2k-4},
\eeq
for $k=2,\cdots, \frac{L-l}{2}-3$, where $c_i:=\cos{\hat{J}_{2i+l+2}}$, $s_i:=\sin{\hat{J}_{2i+l+2}}$, $c':=\cos{B}$, and $s':=\sin{B}$. For fermions near the interfaces,
\beq
    \gamma_{-1}&\longrightarrow c'\gamma_{-1}+c_0 s'\gamma_0- s_0 s'\gamma_3,\\
    \gamma_{0}&\longrightarrow -s'\gamma_{-1}+c_0 c'\gamma_0-s_0 c'\gamma_3,\\
    \gamma_{1}&\longrightarrow c'\gamma_{1}+c_1 s'\gamma_2-s_1 s'\gamma_5,\\
    \gamma_{2}&\longrightarrow -s'\gamma_{1}+c_1 c'\gamma_2- s_1 c'\gamma_5,\\
    \gamma_{L-l-5}&\longrightarrow c_{\frac{L-l}{2}-4}c'\gamma_{L-l-5}+s_{\frac{L-l}{2}-4}c'\gamma_{L-l-8}+s'\gamma_{L-l-4},\\
    \gamma_{L-l-4}&\longrightarrow c'\gamma_{L-l-4}-c_{\frac{L-l}{2}-4}s'\gamma_{L-l-5}-s_{\frac{L-l}{2}-4}s'\gamma_{L-l-8},\\
    \gamma_{L-l-3}&\longrightarrow c_{\frac{L-l}{2}-3}c'\gamma_{L-l-3}+s_{\frac{L-l}{2}-3}c'\gamma_{L-l-6}+s'\gamma_{L-l-2},\\
    \gamma_{L-l-2}&\longrightarrow c'\gamma_{L-l-2}-c_{\frac{L-l}{2}-3}s'\gamma_{L-l-3}-s_{\frac{L-l}{2}-3}s'\gamma_{L-l-6}.
\eeq

To conclude, the Floquet drive corresponds to the following linear map for the fermions 
\beq
    F \gamma_i F^{\dagger} = \sum_j M_{ji} \gamma_j, \,\, M = M'_1 \oplus M_2'
\eeq
where choosing $L-l=20$ for simplicity, we have
\beq
M'_1(\{\hat{J}_k\})=\begin{pmatrix}
c'      & -s'       & 0         & 0         & 0 & 0 & 0 & 0 & 0 & 0  \\
c_0 s'  & c_0 c'    & s_0 c'   & -s_0 s'    & 0 & 0 & 0 & 0 & 0 & 0  \\
-s_0 s'  & -s_0 c'    & c_0 c'    & -c_0 s'   & 0 & 0 & 0 & 0 & 0 & 0   \\
0       & 0         & c_2 s'    & c_2 c'    & s_2 c' & -s_2 s' & 0 & 0 & 0 & 0   \\
0       & 0         & -s_2 s'    & -s_2 c'    & c_2 c' & -c_2 s' & 0 & 0 & 0 & 0  \\
0       & 0         & 0         & 0         & c_4 s' & c_4 c' & s_{4} c' & -s_{4} s' & 0 & 0   \\
0       & 0         & 0         & 0         & -s_4 s' & -s_4 c' & c_{4} c'  & -c_{4} s' & 0 & 0   \\
0       & 0         & 0         & 0         & 0      & 0      & c_{6} s' & c_{6} c' & s_{6} c' & -s_{6} s'   \\
0       & 0         & 0         & 0         & 0      & 0      & -s_{6} s'  & -s_{6} c' & c_{6} c' & -c_{6} s'   \\
0       & 0         & 0         & 0         & 0      & 0      & 0       & 0 & s' & c'   
\end{pmatrix},
\eeq
and 
\beq
M'_2(\{\hat{J}_k\})=\begin{pmatrix}
c'      & -s'       & 0         & 0         & 0 & 0 & 0 & 0 & 0 & 0  \\
c_1 s'  & c_1 c'    & s_1 c'   & -s_1 s'    & 0 & 0 & 0 & 0 & 0 & 0  \\
-s_1 s'  & -s_1 c'    & c_1 c'    & -c_1 s'   & 0 & 0 & 0 & 0 & 0 & 0   \\
0       & 0         & c_3 s'    & c_3 c'    & s_3 c' & -s_3 s' & 0 & 0 & 0 & 0   \\
0       & 0         & -s_3 s'    & -s_3 c'    & c_3 c' & -c_3 s' & 0 & 0 & 0 & 0  \\
0       & 0         & 0         & 0         & c_5 s' & c_5 c' & s_{5} c' & -s_{5} s' & 0 & 0   \\
0       & 0         & 0         & 0         & -s_5 s' & -s_5 c' & c_{5} c'  & -c_{5} s' & 0 & 0   \\
0       & 0         & 0         & 0         & 0      & 0      & c_{7} s' & c_{7} c' & s_{7} c' & -s_{7} s'   \\
0       & 0         & 0         & 0         & 0      & 0      & -s_{7} s'  & -s_{7} c' & c_{7} c' & -c_{7} s'   \\
0       & 0         & 0         & 0         & 0      & 0      & 0       & 0 & s' & c'   
\end{pmatrix}.
\eeq
$M_1'$ has even indices in $c_i,s_i$ and comprises of $\gamma_{4n-1},\gamma_{4n}, n= 0,1 \ldots (L-l)/4 -1$ while $M_2'$ has odd indices in $c_i,s_i$ and comprises of $\gamma_{4n+1},\gamma_{4n+2}, n= 0,1 \ldots (L-l)/4 -1$.
A zero mode 
\beq
    \Psi\propto \sum_{n=0} \Psi_{2n+1}\gamma_{4n-1}+\Psi_{2n+2}\gamma_{4n}. 
\eeq
corresponds to a $+1$-eigenvector of the matrix $M'_1$ (or $M'_2$). We first calculate such an eigenvector for $M'_1$. We have $\Psi_2=-\tan \frac{B}{2}\Psi_1$, and a recursion relation for $k \geq 1$

\beq
    \Psi_{2k}&=c_{2k-2}s'\Psi_{2k-1}+c_{2k-2}c' \Psi_{2k}+s_{2k-2}c' \Psi_{2k+1}-s_{2k-2}s' \Psi_{2k+2},\\
    \Psi_{2k+1}&=-s_{2k-2}s'\Psi_{2k-1}-s_{2k-2}c' \Psi_{2k}+c_{2k-2}c' \Psi_{2k+1}-c_{2k-2}s' \Psi_{2k+2}.
\eeq

Another way to write this recursion relation is
\beq
    \begin{pmatrix}
        \Psi_{2k+1}\\
        \Psi_{2k+2}
    \end{pmatrix}=C_k\begin{pmatrix}
        \Psi_{2k-1}\\
        \Psi_{2k}
    \end{pmatrix},
\eeq
where $C_k$ is the $2\times 2$ matrix
\beq
    C_k&=\frac{1}{s_{2k-2}s'}\begin{pmatrix}
        c_{2k-2}s' & s_{2k-2}s'\\
        -1+c_{2k-2}c' & s_{2k-2}c'
    \end{pmatrix}\cdot \begin{pmatrix}
        -c_{2k-2}s' & 1-c_{2k-2}c'\\
        -s_{2k-2} s' & -s_{2k-2} c'
    \end{pmatrix}\\
    &=\frac{1}{s_{2k-2}s'}\begin{pmatrix}
        -s'^2 & c_{2k-2}s'-s'c'\\
        c_{2k-2}s'-s'c' & -1-c'^2 + 2c_{2k-2}c'
    \end{pmatrix}.
\eeq
\begin{enumerate}
    \item When $\hat{J}_{2(2k-1)+l+2}=J$, this matrix has eigenvectors 
\beq
    \begin{pmatrix}
        1\\
        \cot{B/2}
    \end{pmatrix}\text{ and }\begin{pmatrix}
        1\\
        -\tan{B/2}
    \end{pmatrix},
\eeq
with eigenvalues $-\cot{B/2}\tan{J/2}$ and $-\tan{B/2}\cot{J/2}$, respectively. 
    \item When $\hat{J}_{2(2k-1)+l+2}=0$, we obtain constraints
\beq
    \Psi_{2k}=\cot{\frac{B}{2}}\Psi_{2k-1},
\eeq
and
\beq
    \Psi_{2k+2}=-\tan{\frac{B}{2}}\Psi_{2k+1}.
\eeq
\end{enumerate}

As a special case, when $\hat{J}_n\equiv J$, and taking the $L-l\rightarrow\infty$ limit, the zero modes are given by a solution of the above recursion relation. The zero modes are
\beq
    \Psi&\propto \sum_{n\geq 0}\left(-\frac{\tan \frac{B}{2}}{\tan \frac{J}{2}}\right)^n\left(\gamma_{4n-1}-\tan \frac{B}{2}\gamma_{4n}\right),\\
&=Z_{l+1}- \tan\frac{B}{2} Y_{l+1} + \biggl(-\frac{\tan \frac{B}{2}}{\tan \frac{J}{2}}\biggr)X_{l+1}X_{l+3} \biggl(Z_{l+5}-\tan\frac{B}{2} Y_{l+5}\biggr) + \cdots\\
    \Psi'&\propto \sum_{n\geq 0}\left(-\frac{\tan \frac{B}{2}}{\tan \frac{J}{2}}\right)^n\left(\gamma_{4n+1}-\tan \frac{B}{2}\gamma_{4n+2}\right)\\
    & = X_{l+1} \biggl(Z_{l+3} - \tan\frac{B}{2} Y_{l+3}\biggr) +\left(-\frac{\tan \frac{B}{2}}{\tan \frac{J}{2}}\right) X_{l+1}X_{l+3}X_{l+5}\biggl(Z_{l+7} - \tan\frac{B}{2}Y_{l+7}\biggr) + \cdots
\eeq
When $B<J$, these two zero modes are localized near site $l+1$. When $B=0$, they correspond to $Z_{l+1}$ and $X_{l+1} Z_{l+3}$.

We now discuss the effect of an interaction term $- \frac{U}{2} \sum_i X_{2i}X_{2i+2}$ such that the Hamiltonian is
\beq
    H_{\text{int}}&=H- \frac{U}{2} \sum_i X_{2i}X_{2i+2},\,\, 
    H = \frac{J}{2}H_{\rm trivial|odd} + \frac{B}{2} H_{\rm trivial}
    \label{eq:H_tri_odd_interact}
\eeq
Such an interaction term does not change the SPT nature of the ground state. Here, we study the effect of this interaction out of equilibrium. This interaction commutes with the original Hamiltonian, thus it only changes the energy of each eigenstate. This modifies the Boltzmann weights of the $\{X_{2i}\}$ sectors. For $U/J, U/B \gg 1$, it is easier to just keep track of the number of domain walls on the even sublattice. This gives the following analytic expression for the autocorrelation for the edge mode 
\beq
    \mathcal{A}(t) = (1-q)^{L/2}\cos{\biggl(\frac{B}{J}\biggr)^{L/2} B t} + q(1-q)^{L/2-1}\cos{\biggl(\frac{B}{J}\biggr)^{L/2 -1} B t}+ q^2(1-q)^{L/2-2}\cos{\biggl(\frac{B}{J}\biggr)^{L/2 -2} B t} +\cdots,
    \label{eq:autoco-temperature-interact}
\eeq
where $q\equiv \frac{e^{-\beta U}}{1+e^{-\beta U}}$. Clearly, this boundary operator continues to be localized with interactions, i.e., is a zero mode.

{\bf Floquet dynamics in the $\mathbb{Z}_2^2$ symmetric sector and $\pi$ modes:} Since the $KT$ operator annihilates all the states outside of the $\mathbb{Z}_2^2$ symmetric sector, while behaving as a unitary inside the $\mathbb{Z}_2^2$  symmetric sector, we can study unitary Floquet dynamics within this subspace. While $\Psi,\Psi'$ are not individually  $\mathbb{Z}_2^2$ symmetric, their product is in fact $\mathbb{Z}_2^2$ symmetric. We therefore consider this object, 
\beq
    \Psi_0&=\Psi\Psi'\\
    &=\sum_{n}\left(-\frac{\tan \frac{B}{2}}{\tan \frac{J}{2}}\right)^n \biggl[\sum_{k=0}^n (\prod_{j=0}^{2k-1}X_{2j+l+1})(Z_{4k+l+1}-\tan \frac{B}{2} Y_{4k+l+1})\\
    &\quad (\prod_{j=0}^{2n-2k}X_{2j+l+1})(Z_{4n-4k+l+3}-\tan \frac{B}{2} Y_{4n-4k+l+3})\biggr]
\eeq
If we consider the $KT$ transformation of the above symmetric zero mode term by term, each transformed term is obtained by attaching some Pauli-Z operators on even sites. In particular,
\beq
   KT &\Psi_0 = KT\biggl[\left(Z_{l+1}-\tan \frac{B}{2} Y_{l+1}\right)X_{l+1}\left(Z_{l+3}-\tan \frac{B}{2} Y_{l+3}\right)\\
   &\quad +\left(-\frac{\tan \frac{B}{2}}{\tan \frac{J}{2}}\right)\left(Z_{l+1}-\tan \frac{B}{2} Y_{l+1}\right)X_{l+1}X_{l+3}X_{l+5}\left(Z_{l+7}-\tan \frac{B}{2} Y_{l+7}\right)\\
    &\quad +\left(-\frac{\tan \frac{B}{2}}{\tan \frac{J}{2}}\right)X_{l+1}X_{l+3}\left(Z_{l+5}-\tan \frac{B}{2} Y_{l+5}\right)X_{l+1}\left(Z_{l+3}-\tan \frac{B}{2} Y_{l+3}\right)+\cdots\biggr] \\
    &\quad =\biggl[ X_{l+2}\left(Z_{l+1}-\tan \frac{B}{2} Y_{l+1}\right)X_{l+1}\left(Z_{l+3}-\tan \frac{B}{2} Y_{l+3}\right)\\
    &\quad +\left(-\frac{\tan \frac{B}{2}}{\tan \frac{J}{2}}\right)X_{l+2}X_{l+4}X_{l+6}\left(Z_{l+1}-\tan \frac{B}{2} Y_{l+1}\right)X_{l+1}X_{l+3}X_{l+5}\left(Z_{l+7}-\tan \frac{B}{2} Y_{l+7}\right)\\
    &\quad +\left(-\frac{\tan \frac{B}{2}}{\tan \frac{J}{2}}\right)X_{l+4}X_{l+1}X_{l+3}\left(Z_{l+5}-\tan \frac{B}{2} Y_{l+5}\right)X_{l+1}\left(Z_{l+3}-\tan \frac{B}{2} Y_{l+3}\right)+\cdots\biggr] KT
\eeq
The operator $\Psi_0$ is a zero mode only when we restrict ourself to the sectors of Hilbert space where $\hat{J}_n\equiv J$, which means the Pauli-X operators on even sites (colored in blue above) are simultaneously $-1$ (or $+1$). Therefore, this symmetric zero mode anti-commutes with $KT$ when $X_{2i}=-1$ or commutes with $KT$ when $X_{2i}=1$.

We now plan to include the symmetry operator $KT$ in the Floquet drive. Since $KT$ is non-invertible, the Floquet drive is no longer unitary. However, in the $\mathbb{Z}_2^2$ subspace, the Floquet drive is unitary.
For such a Floquet drive that includes the $KT$ transformation, the approximate zero $\Psi_0$ mode becomes an approximate $\pi$ mode when the Boltzmann weights for the $X_{2i}=-1$ sector dominate over the Boltzmann weights for the $X_{2i}=1$ sector. Fig.~\ref{fig:0-pi-U} shows the autocorrelation function in the presence of the interaction $U$, for two different temperatures, and with and without the $KT$ term as part of the unitary dynamics. The former generates $\pi$ mode dynamics while the latter generates the zero mode dynamics. It is interesting to note that the edge modes show slow oscillating dynamics rather than exponentially decay dynamics, despite many different $\{S_{2i}\}$ sectors participating in the dynamics. 
\begin{figure}[h!]
         \includegraphics[width=0.35\textwidth]{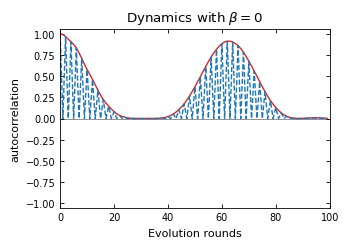}
        \includegraphics[width=0.35\textwidth]{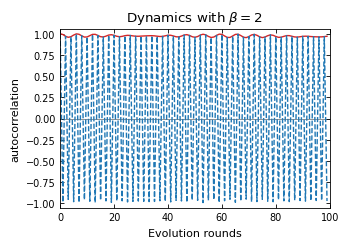}
        \caption{The autocorrelation function of $Y_1 Z_3$ on a chain of size $L=12$, with $J=1$, $B=0.1$ and $U=1$. Red curve represents the dynamics $\langle e^{-\beta H'_{\text{int}}}(F^{\dagger})^n\mathcal{O}F^n\mathcal{O}^{\dagger}\rangle$ with $H'_{\text{int}}=\sum_i (J-B)X_{2i}/2- U X_{2i}X_{2i+2}/2$, while the blue curve represents the dynamics with a $KT$ transformation in each round, i.e. $\langle e^{-\beta H'_{\text{int}}}(F^{\dagger}KT)^n\mathcal{O}(KT F)^n\mathcal{O}^{\dagger}\rangle$. The expectation value is taken in the $\mathbb{Z}_2^2$ symmetric subspace. 
        Left (right) figure shows the result for inverse temperature $\beta=0$ ($\beta=2$).} \label{fig:0-pi-U}
\end{figure}

\end{widetext}

\bibliography{ref}
\end{document}